\documentclass[manuscript]{acmart}

\AtBeginDocument{%
  }

\usepackage{graphicx}
\usepackage{subfigure}
\usepackage{textcomp}
\usepackage{xcolor}
\usepackage{tabu}  
\usepackage{tcolorbox}
\usepackage{booktabs}
\usepackage{multirow}
\tcbuselibrary{hooks}
\usepackage{pifont}
\usepackage[normalem]{ulem}

\setcopyright{acmlicensed}
\copyrightyear{2025}
\acmYear{2025}
\acmDOI{XXXXXXX.XXXXXXX}
\acmISBN{978-1-4503-XXXX-X/2018/06}

\begin{document}

\title{Mind the Gap: A Decade-Scale Empirical Study of Multi-Stakeholder Dynamics in VR Ecosystem}


\author{Yijun Lu}

\affiliation{%
  \institution{Waseda University}
  \city{Tokyo}
  \country{Japan}
}
\email{yijun@ruri.waseda.jp}

\author{Hironori Washizaki}
\authornotemark[1]
\affiliation{%
  \institution{Waseda University}
  \city{Tokyo}
  \country{Japan}
}
\email{washizaki@waseda.jp}

\author{Naoyasu Ubayashi}
\affiliation{%
  \institution{Waseda University}
  \city{Tokyo}
  \country{Japan}
}
\email{ubayashi@aoni.waseda.jp}

\author{Nobukazu Yoshioka}
\affiliation{%
  \institution{Waseda University}
  \city{Tokyo}
  \country{Japan}
}
\email{nobukazu@engineerable.ai}

\author{Chenhao Wu}
\affiliation{%
  \institution{Waseda University}
  \city{Tokyo}
  \country{Japan}
}
\email{wuchenhao@toki.waseda.jp}

\author{Masanari Kondo}
\affiliation{%
  \institution{Kyushu University}
  \city{Kyushu}
  \country{Japan}
}
\email{kondo@ait.kyushu-u.ac.jp}

\author{Yuyin Ma}
\affiliation{%
  \institution{Xinjiang University}
  \city{Urumqi}
  \country{China}
}
\email{mayuyin@xju.edu.cn}

\author{Jiong Dong}
\affiliation{%
  \institution{Xuchang University}
  \city{Xuchang}
  \country{China}
}
\email{jiongdong@xcu.edu.cn}

\author{Jianjin Zhao}
\affiliation{%
  \institution{Beijing University of Posts and Telecommunications}
  \city{Beijing}
  \country{China}
}
\email{jianjinzhao@bupt.edu.cn}

\author{Dongqi Han}
\affiliation{%
  \institution{Beijing University of Posts and Telecommunications}
  \city{Beijing}
  \country{China}
}
\email{handongqi@bupt.edu.cn}

\renewcommand{\shortauthors}{Lu and Washizaki et al.}

\begin{abstract}

In the development and evolution of VR ecosystem, platform stakeholders continuously adapt their products in response to user and technical feedback, often reflected in subtle shifts in discussion topics or system updates. A comprehensive understanding of these changes is essential for identifying gaps between user expectations and developer actions, which can guide more effective quality assurance and user-centered innovation. While previous studies have analyzed either user reviews or developer discussions in isolation, such approaches typically fail to reveal how specific user concerns are (or are not) addressed by corresponding technical activities. To address this limitation, our study introduces a multi-view empirical framework that systematically compares and aligns stakeholder perspectives. By applying topic modeling and quantitative impact analysis to 944,320 user reviews and 389,477 developer posts, we identify not only the overlap in concerns (e.g., performance, input methods), but also clear gaps in areas like inclusivity and community safety (e.g., LGBTQ+ representation, child-friendly content). Our findings show that while users repeatedly raise such issues, they are rarely discussed in developer forums. These insights enable data-driven recommendations for closing the user-developer gap in VR ecosystems, offering practical implications for platform governance and the design of next-generation VR systems.

\end{abstract}




\maketitle

\section{Introduction}

Virtual Reality (VR) has rapidly evolved from a niche technology into a widely adopted medium, fundamentally transforming both the ways users interact with software and how developers approach application design. The global VR games market was valued at \$47.1 billion in 2024 and is projected to surge to \$346.0 billion by 2033~\cite{market}. This remarkable expansion is driven by increasingly immersive user experiences, which generate rich streams of user feedback that illuminate VR-specific needs and usability challenges~\cite{epp2021empirical}. Understanding these nuanced user perspectives is essential, as VR players frequently express concerns (e.g., motion sickness, inclusivity, and hardware limitations) that are less prevalent in traditional software~\cite{lu2024empirical}. Addressing these unique issues is critical for developers seeking to align products with evolving user expectations and to fully realize the potential of VR platforms~\cite{lu2024empirical}. At the same time, the VR ecosystem involves a diverse set of stakeholders, from independent developers to global studios, whose priorities, constraints, and interpretations often diverge~\cite{rodriguez2021perceived}. Bridging the persistent gap between user experiences and developer practices is therefore central to advancing VR technology.

However, the current research landscape is dominated by single-stakeholder analytics, with most prior studies focusing exclusively on either user feedback or developer discourse. On the user side, analyses of app store reviews (including those for VR applications) have revealed user preferences and criticisms, most notably frequent requests for enhanced immersion, richer content, and better accessibility~\cite{Khalid, dong2024user}. For example, Dong et al.\cite{dong2024user} found that while users value aspects like music and gameplay, their most persistent complaints concern bugs, insufficient content, and high costs. On the developer side, empirical analyses of Q\&A forums, code repositories, and technical discussions have revealed the practical challenges that developers encounter, including toolchain adaptation, cross-platform support, and performance optimization\cite{barua2014stackoverflow}. 
Although each of these research streams offers important perspectives, their separation makes it challenging to trace the translation of user concerns into developer priorities and technical solutions.

This separation has led to a well-documented misalignment between user needs and developer focus, especially in rapidly evolving domains like VR. As recent work shows, user feedback in VR shifts quickly from foundational concerns such as physical comfort to higher-level expectations involving social interaction, content diversity, and community safety~\cite{epp2021empirical, lu2024empirical}. However, developers may not adapt with these changes, resulting in critical issues such as inclusivity, accessibility, and safety receiving insufficient attention in development processes. For example, our empirical results show that inclusivity topics, e.g., the representation of LGBTQ+ users or child-friendly content, are discussed by users but receive minimal attention in developer forums. Conversely, developers devote considerable discussion to technical challenges that may not directly address user frustrations or desires. This persistent gap underscores the necessity of integrated, multi-stakeholder empirical analysis in VR ecosystems.

\begin{figure*}[t]
	\centering
	\includegraphics[width=\textwidth]{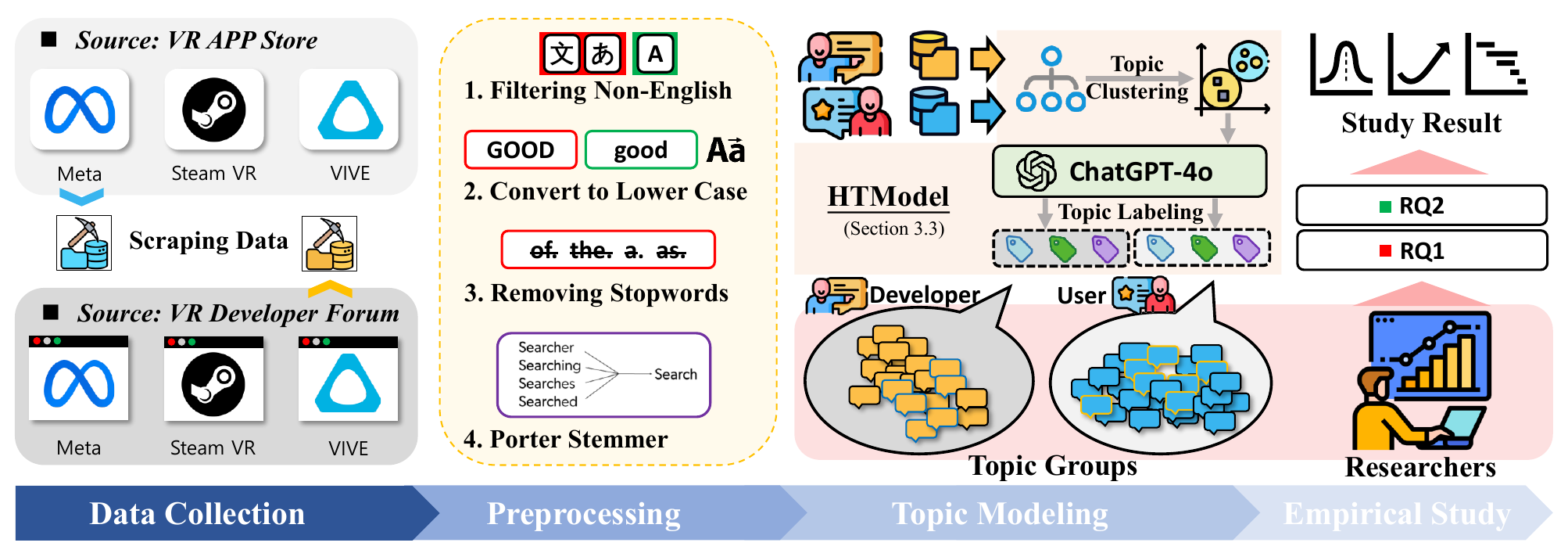}
	\caption{Overview of the research workflow, encompassing data collection from major VR platforms, comprehensive preprocessing, a hybrid topic modeling pipeline, and empirical analysis. This workflow enables systematic investigation of user–developer dynamics in the VR ecosystem by integrating large-scale data acquisition, standardized text processing, advanced topic extraction, and multi-view analysis.}
	\label{Overflow}
\end{figure*}

To address this critical gap, we conduct the first large-scale empirical study that systematically integrates and compares the perspectives of both VR users and developers. Leveraging a dataset of 944,320 user reviews and 389,477 developer posts from three major VR platforms (Mate, SteamVR, and VivePort), we provide a comprehensive view of stakeholder dialogue and its evolution over time. For efficient analysis of this extensive corpus, we introduce a hybrid approach, HTModel (See Section \ref{TopicExtraction}), which decouples topic discovery from semantic interpretation by combining scalable clustering with LLM-based labeling to produce interpretable topic maps. We perform topic modeling separately on user and developer datasets, then merge and categorize the resulting topics into stakeholder-relevant themes. This enables absolute and relative impact analysis for each topic across both groups. Our unified framework identifies overlooked user concerns, highlights stakeholder alignment, and provides actionable insights for research and practice. In summary, this work fills a key gap in VR analytics by offering a reproducible, multi-view framework that links user experience with developer challenges. The workflow is summarized in Fig. \ref{Overflow}.

Specifically, we address the following research questions (RQs), along with a summary of the key findings for each. Detailed results are presented in Section \ref{emprical}.

\begin{itemize}
    \item \textbf{RQ1: What topics are most frequently discussed by VR developers and users?}

\textit{Motivation:} This research question aims to reveal the most prominent topics discussed by each stakeholder group and highlight key mismatches in their priorities.

\textit{Results:} Technical performance issues such as hardware support (users 2.6\%, developers 2.1\%) and input methods (users 1.8\%, developers 2.8\%) are common to both groups. In contrast, inclusivity and community-related topics including LGBTQ+ representation (users 1.9\%), child-friendly content (users 2.0\%), and female character inclusivity (users 2.4\%) are frequently raised by users but are virtually absent in developer forums. Users also mention language support (1.7\%) and pricing (3.7\%) more often, whereas developers discuss toolchains (4.0\%), cross-platform issues (3.8\%), and server management (2.8\%) more frequently. These findings quantify a clear misalignment, especially regarding social, inclusivity, and accessibility issues.

    \item \textbf{RQ2: How do these topics evolve over time from different perspectives?}

\textit{Motivation:} This research question investigates whether the focus of each stakeholder group shifts over time and how their evolving concerns converge or diverge.

\textit{Results:} Software-related topics accounted for up to 60\% of user discussions in earlier years. Since 2021, mentions of content quality, replay value, and inclusivity have increased by more than 30\%. Community experience, child-friendly design, and diverse avatars have become notably more prominent in recent user reviews. Developer focus on user experience increased to about 20\% by 2024, but inclusivity, safety, and accessibility topics remain almost absent in developer posts, even as these themes appear in up to 6\% of user comments during the same period. Developer attention to new user concerns often lags behind shifts in user discussions.
\end{itemize}

The primary contributions of this study are summarized as follows:

\begin{itemize}
    \item We conduct the first large-scale empirical study integrating and systematically comparing the perspectives of both users and developers in VR ecosystems, analyzing 944,320 user reviews and 389,477 developer posts collected from three major VR platforms.

    \item We identify and quantify significant gaps between user-centric concerns (e.g., inclusivity, accessibility, pricing) and developer-focused priorities (e.g., toolchains, cross-platform compatibility), revealing critical gaps between user expectations and developer practices.

    \item We track the temporal evolution of discussion topics within each stakeholder group, highlighting areas where developer attention lags behind emerging user needs, such as community experience, diverse avatars, and child-friendly content.

    \item To facilitate the analysis, we propose HTModel, a hybrid topic-modeling pipeline combining traditional clustering methods with LLM-based semantic labeling. HTModel significantly reduces manual labeling effort compared to traditional topic models and achieves over 99\% cost savings in LLM API usage compared to purely LLM-driven approaches.

\item We provide the first publicly available multi-stakeholder textual dataset from major VR platforms, supporting future reproducibility and empirical research in the VR ecosystem. The dataset is available \href{https://github.com/asi1117/TOSEM.git}{here}.

\end{itemize}
The remainder of this paper is structured as follows. Section~\ref{related work} reviews the relevant literature. Section~\ref{methodlogy} details the data collection, processing, and topic extraction methodology. Section~\ref{emprical} presents the empirical study of user reviews and developer posts. Section~\ref{implication} discusses the broader implications of our findings. Section~\ref{limitation} outlines threats to validity. Finally, Section~\ref{conclusion} concludes the paper. Additional methodological details are provided in the Appendix.

\section{Related Work}
\label{related work}

Building on this foundation, our work integrates three key research streams: (1) empirical multi-stakeholder analysis in software engineering, (2) empirical studies focused on VR systems, and (3) advances in empirical methods for topic modeling. By synthesizing these perspectives, we address existing limitations and provide a unified framework for comprehensive multi-view analysis in the VR domain.

\subsection{Multi‑stakeholder Empirical Studies in Software Engineering}

Research in software engineering has long emphasized the need to align perspectives across stakeholder groups. For instance, Hassan et al.\cite{hassan2018dialogue} showed that many user-reported issues in Google Play app reviews are never addressed in developer discussions, leading to unresolved bugs and unmet user expectations. Buchan et al.\cite{buchan2021alignment} used repertory grid techniques in agile teams to uncover subtle misalignments between user representatives and developers, illustrating how differing expectations can hinder collaboration. Similarly, Hasan et al.\cite{hasan2021survey} surveyed 181 professional developers and found that demands from different stakeholders, such as peers, managers, and users, substantially impact developer satisfaction and productivity. Lenberg et al.\cite{lenberg2018misaligned} further demonstrated that organizational value misalignment is significantly associated with reduced team performance and increased conflict. Additionally, Mauerer et al.~\cite{mauerer2021socio} conducted a longitudinal study of 25 open-source projects, revealing that misalignment between social structures and technical dependencies does not always result in poor code quality, highlighting the complexity of stakeholder alignment in software ecosystems. Despite this growing body of work, few studies have directly examined the distinct challenges of stakeholder alignment in VR systems.

\subsection{VR Empirical Studies with a Single Stakeholder Focus}

Empirical research in VR has predominantly relied on single-stakeholder perspectives, which provide useful but incomplete insights. On the user side, text-mining studies of app reviews are common. Lu et al.~\cite{lu2024empirical} analyzed over 176,000 VR game reviews, identifying user concerns such as performance, tracking, and compatibility. Dong et al.~\cite{dong2024user} examined more than 105,000 social VR reviews and highlighted the demand for avatar customization. Li et al.~\cite{li2023towards_quality} further analyzed over one million user reviews to develop a taxonomy of VR-specific quality attributes. Additional studies focus on immersive usability: Zhang et al.~\cite{zhang2025cvr_usability} conducted a meta-analysis of Cinematic VR, revealing inconsistent measurement of presence and immersion, while Rzig et al.~\cite{rzig2022vr_testing} found that 74\% of open-source VR projects lacked adequate test coverage. Singh and O’Hagan~\cite{singh2024socialvr} used topic modeling on 40,000 Rec Room reviews, identifying emergent issues such as harassment through sentiment analysis.

On the developer side, separate studies have addressed VR-specific tool and testing challenges. Karre et al.~\cite{karre2019vr_practices} surveyed VR software teams across several countries and found that hybrid engineering practices are needed but understudied. Survey-based work by Ashtari et al.~\cite{ashtari2023arvr_challenges} and Dhiaet al. ~\cite{rzig2022vr_testing} report ongoing deficiencies in testing workflows, tool support, and quality assurance. While these studies enhance our understanding of either user or developer needs in VR, there is still a lack of large-scale, integrated analysis of both perspectives. It remains uncertain how developer priorities correspond with user concerns or how quickly developers respond to shifting user expectations. Addressing these gaps, our work is the first to systematically compare user reviews and developer discussions in the VR domain through a multi-stakeholder lens.

\subsection{Advances in Empirical Methods for Topic Modeling}

Automated methods for topic analysis have evolved significantly, ranging from early semantic approaches like Latent Semantic Analysis and Non‐negative Matrix Factorization (LSA, NMF) to probabilistic techniques such as Latent Dirichlet Allocation (LDA)~\cite{blei2003latent, chang2009reading}. While LDA improves topic discovery, it still suffers from label ambiguity and requires expert interpretation~\cite{ramage2011partially, epp2021empirical}. Embedding-based models such as BERTopic and Top2Vec~\cite{grootendorst2022bertopic, angelov2020top2vec} further enhance coherence and handle short texts more effectively. However, these models retain the need for manual labeling of topic clusters~\cite{ramage2011partially}.

The rise of large language models (LLMs) has enabled the automation of topic labeling. Approaches like TopicGPT~\cite{pham2023topicgpt}, generative semantic labeling techniques~\cite{kozlowski2024generative}, and LLM‐based frameworks such as LimTopic~\cite{azher2024limtopic} and kapoor’s method for BERTopic labels~\cite{kapoor2024qualitative} demonstrate improved interpretability, often matching or surpassing human judgment~\cite{azher2024limtopic, kapoor2024qualitative}. At the same time, iterative methods like LITA~\cite{chang2025lita} show that combining LLMs with traditional clustering can improve topic coherence and reduce labeling effort. Despite these advancements, purely LLM‐driven labeling remains expensive and difficult to scale to millions of documents, as costs grow linearly with data volume~\cite{pham2023topicgpt, kozlowski2024generative}. Our HTModel builds on these foundations by first applying efficient traditional clustering techniques (i.e., LDA, BERTopic, NMF, LSA, and Top2Vec.) at scale, then selectively using LLMs to assign semantic labels to clusters. This hybrid design achieves interpretable, semantically rich topic extraction across large corpora while drastically reducing manual annotation and API costs.

\section{Methodology }
\label{methodlogy}

To support the subsequent VR empirical study and automated analysis, we collect VR-related textual data, apply noise reduction, and introduce a hybrid pipeline that combines traditional topic models with GPT-4o to realize enhanced topic extraction.


\begin{table}[t]
    \centering
    
\caption{Dataset Characteristics. The ${D_{developer}}$ dataset covers the period from 2013 to 2024, and the ${D_{user}}$ dataset spans from 2015 to 2024. All data were collected and finalized between July 1–15, 2024.}

    \begin{tabular}{l l cc}
        \toprule
        \textbf{Dataset} & \textbf{Platforms} & \textbf{Number of Collected Data} & \textbf{Number of Processed Data} \\
        \midrule
        \multirow{3}{*}{${D_{developer}}$}
            & VivePort & 43,053 & 41,782 \\
            & Meta  & 352,463 & 342,104 \\
            & SteamVR  & 5,932 & 5,591 \\
        \cmidrule(lr){2-4}
            & Subtotal & 401,448 & 389,477 \\
        \midrule
        \multirow{3}{*}{${D_{user}}$}
            & VivePort  & 17,844 & 15,964 \\
            & Meta & 393,355 & 351,480 \\
            & SteamVR  & 646,642 & 576,876 \\
        \cmidrule(lr){2-4}
            & Subtotal & 1,057,841 & 944,320 \\
        \midrule
        Total &  & 1,459,289 & 1,333,797 \\
        \bottomrule
    \end{tabular}
    \label{tab:data}
\end{table}

\subsection{Data Collection}
\label{DataCollection}

To enable a comprehensive multi-stakeholder analysis, we collected large-scale textual data from three leading VR gaming platforms: Meta~\cite{Meta2024}, SteamVR~\cite{SteamVR2024}, and Viveport~\cite{Viveport2024}. These platforms are widely recognized in the literature for their extensive VR game catalogs and vibrant user communities, making them representative sources for studying user experience in the VR domain~\cite{lu2024empirical,dong2024points,epp2021empirical}. By focusing on VR\_Only games, we ensured that the collected user reviews directly reflect authentic VR experiences, in line with established empirical research practices~\cite{lu2024empirical, dong2024user}. 

In addition to user reviews, our work uniquely incorporates large-scale developer forum data from the Meta Forum~\cite{MetaCommunity2024}, SteamVR Forum~\cite{SteamCommunityApp2024}, and HTC Vive Forum~\cite{HTCDeveloper2024}. While forum-based analysis has proven valuable for exploring collaboration and technical knowledge in software engineering and open source communities~\cite{kamienski2021empirical,uddin2021empirical}, the systematic study of VR developer forums remains rare. By combining both user and developer sources, our dataset enables a holistic investigation of the dialogue and gaps between end-user expectations and developer practices in the VR ecosystem.

We developed customized web crawlers to systematically collect and synchronize user reviews and developer discussions from these platforms. The resulting datasets, $D_{user}$ and $D_{developer}$, support robust, multi-perspective topic modeling and empirical analysis. See Table~\ref{tab:data} for detailed statistics.

\subsection{Data Processing}
\label{DataProcessing}

To ensure the reliability and consistency of subsequent analyses, both $D_{user}$ and $D_{developer}$ datasets underwent rigorous preprocessing, following best practices in large-scale empirical software analytics~\cite{hardeniya2016natural, mckinney2011pandas, Stahl2020}. The main steps are summarized as follows:

\begin{itemize}
\item \textbf{Filtering Non-English:} Only English-language content was retained using the \textit{lingua-language-detector}\cite{Stahl2020}, as many state-of-the-art natural language processing models are primarily trained and evaluated on English corpora, and thus exhibit optimal performance or applicability for English text\cite{lin2019empirical}.
\item \textbf{Convert to Lowercase:} All textual data were converted to lowercase using Pandas~\cite{mckinney2011pandas}, ensuring case-insensitive processing and vocabulary normalization, thereby reducing data sparsity and potential analytical artifacts in downstream modeling.
\item \textbf{Removing Stopwords:} Standard stopword removal, including the application of a domain-specific (VR-related) stopword list, was conducted with NLTK~\cite{hardeniya2016natural}. This step serves to eliminate high-frequency, semantically uninformative terms, thus enhancing the interpretability and coherence of extracted topics.
\item \textbf{Porter Stemmer:} The Porter stemming algorithm~\cite{permana2021stemming} was employed to standardize inflected word forms to their roots, reduce feature dimensionality, and improve the effectiveness of topic modeling and other downstream natural language processing tasks~\cite{ramage2011partially, lin2019empirical}.
\end{itemize}

These steps collectively ensure high-quality input for subsequent topic extraction. The final dataset statistics after processing are shown in Table~\ref{tab:data}.

\begin{figure}[t]
    \centering
    \includegraphics[width=0.95\columnwidth]{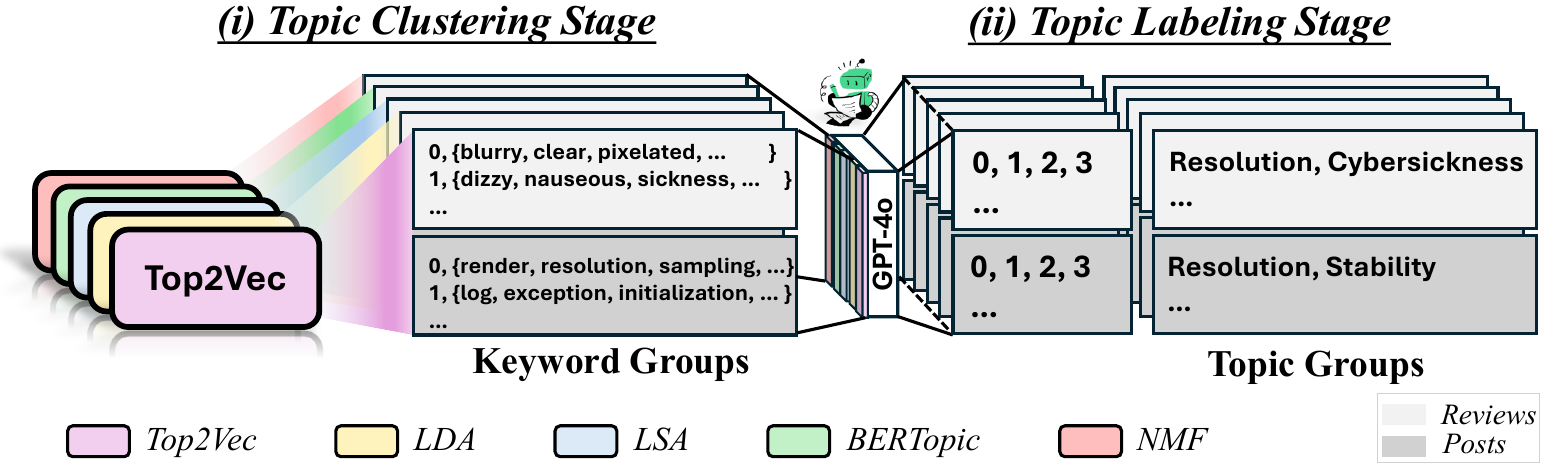}
\caption{Implementation process of the proposed Hybrid Topic Modeling (HTModel). The pipeline comprises two main stages: (i) parallel topic clustering using multiple algorithms (Top2Vec, LDA, LSA, BERTopic, and NMF) to generate diverse keyword groups, and (ii) automated topic labeling with ChatGPT-4o to produce unified, interpretable topic groups for downstream analysis.}
    \label{HTModel}
\end{figure}

\subsection{Topic Extraction}
\label{TopicExtraction}

To achieve robust and interpretable topic analysis for multi-stakeholder VR textual data, we propose the HTModel pipeline to extract topics. First, we employ five topic modeling approaches, namely LDA, LSA, NMF, BERTopic, and Top2Vec, to cluster large-scale textual data effectively, thereby reducing reliance on computationally intensive LLM APIs. We then leverage GPT-4o to enhance topic interpretability by generating semantically refined labels for the extracted keyword clusters, thus eliminating the need for manual annotation. The overall workflow of the proposed HTModel, encompassing both topic clustering and topic labeling, is illustrated in Figure~\ref{HTModel}.

\subsubsection{Topic Clustering }

Building on the foundational work of Roman et al.~\cite{egger2022topic}, we select the LDA, BERTopic, Top2Vec, LSA, and NMF topic models to perform a comparative analysis across datasets $D_{developer}$ and $D_{user}$. LDA represents generative models, and the LSA and NMF models are based on matrix decomposition. In contrast, the BERTopic and Top2Vec models employ embedding-based clustering techniques.

These five models are chosen because they collectively represent the three dominant families of topic modeling approaches in recent research: (1) probabilistic models (LDA), (2) matrix factorization models (LSA, NMF), and (3) embedding-based clustering models (BERTopic, Top2Vec). This selection ensures a comprehensive and balanced comparison, capturing the methodological diversity and current state-of-the-art in the topic modeling literature~\cite{egger2022topic, grootendorst2022bertopic, angelov2020top2vec, blei2003latent, chang2009reading}.

To ensure a consistent and fair evaluation, we implement all five topic models with a fixed number of topics ($T = 50$) and maintain the number of keywords per topic at $Word = 10$, aligning with parameters commonly used in manual annotation~\cite{lu2024empirical, dong2024points, epp2021empirical}. This consistency allows for a direct comparison between traditional manual labeling and our GPT-based approach. Note that the topics generated by these models are essentially clusters of keywords that serve as the foundation for subsequent semantic refinement and labeling.

\subsubsection{Topic Labeling }

After applying the five conventional topic models for topic clustering, the large-scale datasets yield 250 representative topics, each represented by a group of 10 related keywords. These keyword groups are then input to GPT-4o to generate the topic labels for each group, a process that replaces traditional manual annotation tasks. Drawing inspiration from recent advances in LLM-based topic modeling ~\cite{pham2023topicgpt, kozlowski2024generative, azher2024limtopic, chang2025lita, kapoor2024qualitative}, we further refine and adapt the prompts to suit the characteristics of our datasets and research objectives.

All experiments utilize OpenAI GPT-4o (API access dates: November 1, 2024, to January 1, 2025), during which the GPT-4o model parameters remain stable, ensuring consistency and reliability across experimental trials. Moreover, we conduct systematic prompt engineering and confirm that minor variations in prompt design have minimal impact on the stability and accuracy of the semantic labeling process.

\begin{tcolorbox}[
    title={Prompt Design for Semantic Labeling of $D_{user}$ and $D_{developer}$},
    colbacktitle=gray!35!white,
    coltitle=black,
    colback=gray!8, 
    colframe=black, 
    width=\linewidth, 
    arc=1mm, auto outer arc,
    boxrule=0.5pt,
    boxsep=1pt, 
    left=2pt, right=2pt, top=2pt, bottom=2pt 
]
\begin{minipage}{0.48\linewidth}
    \textbf{Prompt 1: User Review Topic Labeling} \\[0.5pt]
    This analysis categorizes VR user reviews into 50 topics. One topic is defined by the following keywords: \{keywords\}. Generate a concise, semantically meaningful label that best represents this topic. Use the format: \\
    \textbf{Topic:} <topic label>.
\end{minipage}%
\hfill
\begin{minipage}{0.48\linewidth}
    \textbf{Prompt 2: Developer Post Topic Labeling} \\[0.5pt]
    This analysis categorizes VR developer discussions into 50 topics. One topic is defined by the following keywords: \{keywords\}. Generate a concise, semantically meaningful label that best represents this topic. Use the format: \\
    \textbf{Topic:} <topic label>.
\end{minipage}
\end{tcolorbox}

This two-stage pipeline enables interpretable and scalable topic extraction for millions of documents across both user and developer datasets. We conducted both qualitative and quantitative evaluations of the hybrid model and ultimately selected LDA\_GPT as the analytical framework. Through multiple rounds of sensitivity analysis, we determined the optimal number of topics to be $T_{D_{developer}} = 40$ for developer discussions and $T_{D_{user}} = 50$ for user discussions, with the number of keywords set to $K = 20$. Further methodological details can be found in Section~\ref{Modeling}.

\section{Empirical study}

\label{emprical}



As mentioned previously, a multiview analysis is conducted in this study to identify the key concerns of VR users and the primary priorities of developers. 
Unlike previous studies that focus primarily on a single perspective (e.g., users or developers), the current study integrates both user and developer perspectives to systematically investigate similarities and differences and analyze how these evolve over time in the VR domain. Thus, this study is to explore a more comprehensive understanding of these dynamics by leveraging a dataset of user reviews (denoted with \textcolor{blue}{blue} R-prefixed IDs) and developer posts (denoted with \textcolor{orange}{orange} P-prefixed IDs).

After topic modeling and automated labeling, each comment and post is assigned to a specific topic. As shown in Figures \ref{fig:user_topic_rate} and \ref{fig:developer_topic_rate}, the average number of reviews or posts per topic is 21,582 for the user dataset and 10,246 for the developer dataset. According to standard statistical formulas, the minimum required sample sizes to achieve a 95\% confidence level with a ±5\% margin of error are 378 and 370, respectively. To further enhance the robustness and reliability of our qualitative analysis, we conservatively select 500 reviews or posts per topic. The qualitative analysis is independently conducted by two authors per dataset, each possessing more than five years of development experience and substantial expertise in VR.

\subsection{(RQ1) Which topics are most frequently discussed by VR developers and users}


\subsubsection{Motivation}

User reviews and developer forum discussions serve as critical resources for understanding the VR domain, offering dual perspectives on players’ perceptions of game quality and the technical challenges faced by developers. 

In this study, we explore overlaps and divergences between perspectives of the users and developers by conducting a multiview textual data analysis, aiming to bridge the experience-implementation gap between user feedback and developer challenges.
The results of this analysis provide actionable insights for designing VR systems that better align with the user's expectations while also addressing the real-world technical and practical constraints of developers.

\subsubsection{Approach}

In Section~\ref{Selection}, we evaluate the effectiveness of the proposed HTModel and select the LDA\_GPT model for topic extraction. Based on the scale and content of the datasets, we determine the optimal number of topics using the coherence score, resulting in $T_{D_{developer}} = 40$ topics and $T_{D_{user}} = 50$ topics for the developer and user datasets, respectively. 

\begin{figure*}[t]
	\centering
	\includegraphics[width=\textwidth]{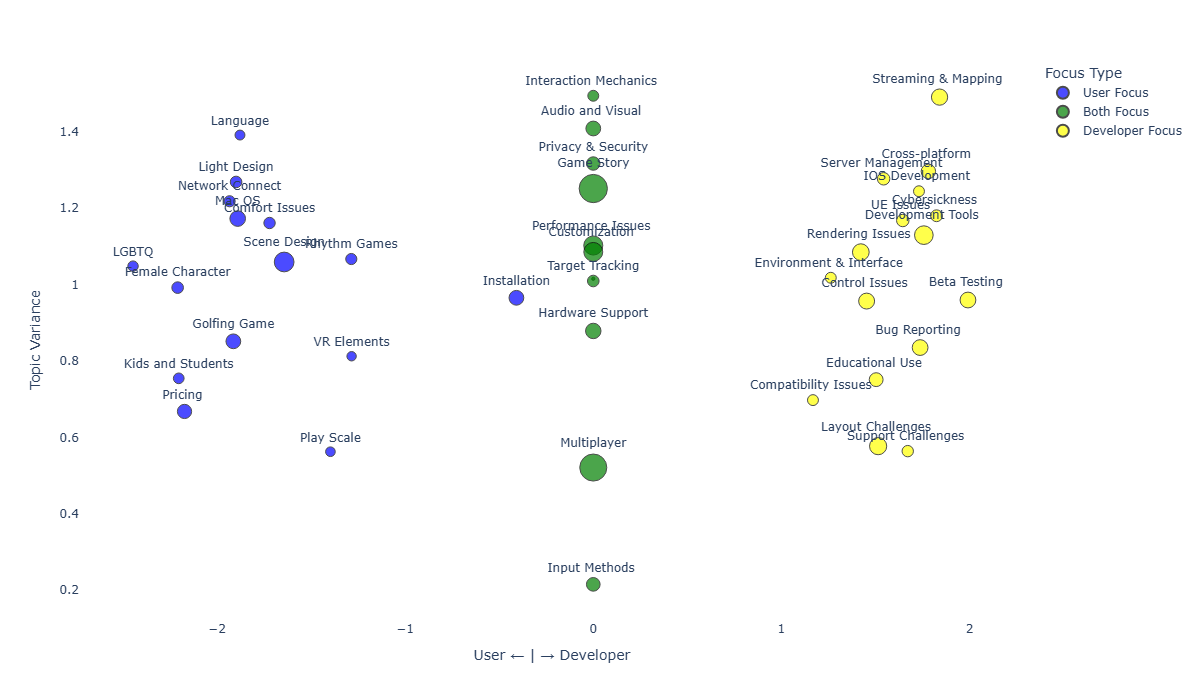}
\caption{Topic distribution after automated merging using S-BERT semantic analysis and cosine similarity. 
Blue nodes indicate user-focused topics, yellow indicate developer-focused topics, and green indicate topics of mutual interest. 
The horizontal axis reflects relative focus (left: user-oriented, right: developer-oriented), and the vertical axis shows variance in discussion prominence between the two groups. 
Positions are derived from S-BERT embeddings projected via t-SNE.}
	\label{TopicOverlap}
\end{figure*}

To further analyze the similarities and differences in focus between developers and users, a two-step method is proposed to calculate the intra-dataset and the inter-dataset topic similarity. Specifically, we employ the Sentence-BERT (S-BERT) model~\cite{reimers2019sentence} and cosine similarity to address the potential redundancy in topic labels in the first step. We then calculate the inter-dataset topic similarity, the cosine similarity is shown as follows:



\begin{equation}
CoS_{A_i, B_i} = \frac{\sum_{i=1}^{n} A_i \cdot B_i}{\sqrt{\sum_{i=1}^{n} A_i^2} \cdot \sqrt{\sum_{i=1}^{n} B_i^2}}
\end{equation}  
where $A_i$ and $B_i$ denote the feature vectors of the two topic labels. We conduct a sensitivity analysis to determine the optimal merging threshold. When the threshold is set too low (e.g., 0.7), semantically distinct topics such as \textit{Bug Reporting} and \textit{Troubleshooting} are erroneously merged, resulting in excessive generalization. Conversely, a high threshold (e.g., 0.9) prevents the merging of highly similar topics, such as \textit{Graphics Optimization} and \textit{Rendering Techniques}, leading to fragmented and redundant labels. Through empirical validation, we set the merging threshold to $0.8$, which achieves the best trade-off between topic coherence and granularity. 
After the first step, we extract $T_{D_{user}} = 43$ and $T_{D_{developer}} = 37$ topics from user and developer datasets, respectively. That is, $T_{D_{user}} = 43$ and $T_{D_{developer}} = 37$ topics are assigned to associated documents, aiming to quantify the document count for each discussion topic. 
We then obtain the similarities and differences between developers and users, as shown in Figure~\ref{TopicOverlap}.


\subsubsection{Results}

\begin{figure}[t]
    \centering
    \includegraphics[width=\textwidth]{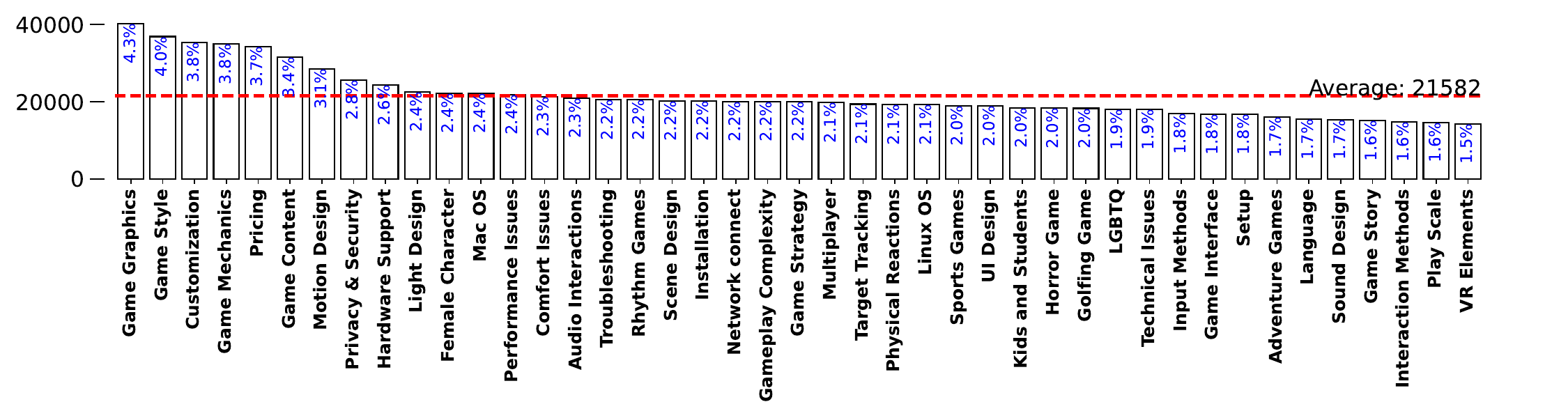}
\caption{Topic distribution in $D_{user}$ (43 extracted topics). The X-axis represents the topics, and the Y-axis indicates the corresponding number of user reviews. The red line denotes the average number of reviews across all topics, while the blue percentages indicate the proportion of reviews for each topic.}

    \label{fig:user_topic_rate}
\end{figure}

\begin{figure}[t]
    \centering
    \includegraphics[width=\textwidth]{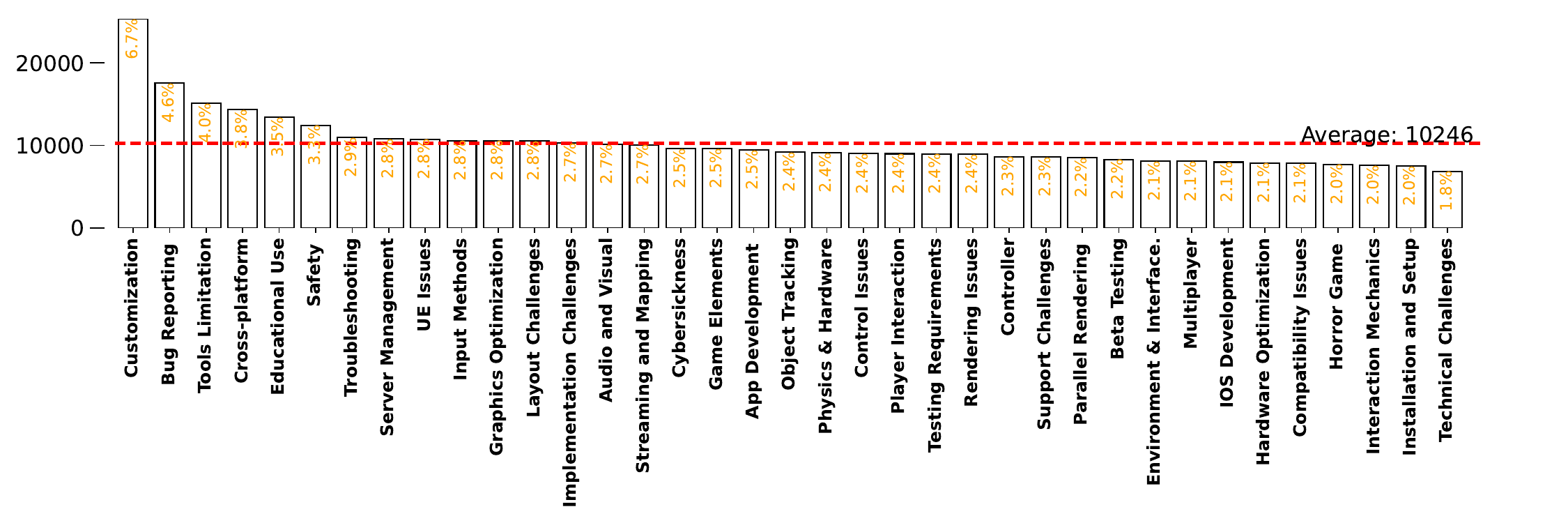}
\caption{Topic distribution in $D_{developer}$ (37 extracted topics). The X-axis represents the topics, and the Y-axis indicates the corresponding number of developer posts. The red line denotes the average number of posts across all topics, while the orange percentages indicate the proportion of posts for each topic.}

    \label{fig:developer_topic_rate}
\end{figure}

Figure \ref{fig:user_topic_rate} shows the 43 topics extracted from the user reviews ranked by review volume. On average, each topic includes 21,582 reviews. Notably, 13 topics exceed this average, indicating key areas of user interest and concern. In addition, Figure \ref{fig:developer_topic_rate} shows the distribution of the 37 topics extracted from developer forum discussions ranked by post volume. On average, each topic includes 10,246 discussions. Notably, 12 topics exceed this average, highlighting key areas of developer focus and technical challenges.

Figure \ref{TopicOverlap} displays the shared and distinct focus areas of users and developers, where green circles indicate overlapping topics between game reviews and forum discussions, blue circles represent topics unique to reviews, and orange circles denote those exclusive to forums.

\textbf{Overlapping Focus.} The analysis revealed distinct yet overlapping priorities between the users and developers. For example, in terms of \textbf{input methods} (1.8\% \& 2.8\%), users appreciate support for diverse peripherals, e.g., hands on throttle-and-stick HOTAS devices, mouse and keyboard, controllers, and Vive wands (\textcolor{blue}{R580062}), while developers focus on addressing technical challenges and proposing workarounds for input-related issues (\textcolor{orange}{P1182045}). In terms of \textbf{multiplayer} considerations (2.1\% \& 2.1\%), users advocate for more cooperative experiences (\textcolor{blue}{R139393639}), and developers focus on troubleshooting implementation errors (\textcolor{orange}{P750531}).

For \textbf{hardware support} (2.6\% \& 2.1\%), users emphasize the need for broader compatibility, including specialized peripherals, e.g., flight hardware (\textcolor{blue}{R670366}), and developers highlight the impact of hardware compatibility on development efficiency (\textcolor{orange}{P96008}). In addition, in terms of \textbf{target tracking} (2.1\% \& 2.4\%), users report issues associated with tracking specific body parts, e.g., hands and feet (\textcolor{blue}{R705224}), whereas developers analyze factors like LED positioning to improve accuracy (\textcolor{orange}{P140481}). Regarding \textbf{customization} (3.8\% \& 6.7\%), users express dissatisfaction with limited options (\textcolor{blue}{R581554}), and developers investigate strategies to enhance personalization, e.g., defining hairstyles through arrays (\textcolor{orange}{P153285}).

\textbf{Performance} (2.4\% \& 2.8\%) was identified as a recurring concern. Users acknowledge that initial issues have been resolved through updates (\textcolor{blue}{R157371048}), and developers propose optimizations, e.g., tailoring speaker placement based on head profiles (\textcolor{orange}{P68101}). In addition, \textbf{game elements} (2.4\% \& 2.5\%) are critical in terms of player retention, with users praising immersive narratives (\textcolor{blue}{R544047}), and developers focusing on interaction-driven storytelling over more conventional mechanics (\textcolor{orange}{P109287}).

\textbf{Privacy and security} (2.8\% \& 3.3\%) are also significant concerns. Users request private spaces in games like \textit{Star Trek VR} (\textcolor{blue}{R204994}) and raise alarms about excessive data collection (\textcolor{blue}{R119697824}). Developers address these concerns by restricting raw camera data access and providing alternative methods, e.g., the Wave 5.1.0 SDK, for application development (\textcolor{orange}{P50679}). In addition, \textbf{audio and visual elements} (8.1\% \& 2.7\%) are pivotal for user immersion (\textcolor{blue}{R65580}), while developers explore integrating modern music into VR environments (\textcolor{orange}{P557552}). In terms of \textbf{interaction elements} (3.1\% \& 4.4\%), well-designed mechanics in games like \textit{Half-Life: Alyx}, \textit{Lone Echo 1}, and \textit{S.T.A.L.K.E.R} enhance engagement considerably (\textcolor{blue}{R516719}, \textcolor{blue}{R337390}, \textcolor{blue}{R364171}), with developers striving to improve the quality of interactions through innovations, e.g., gesture-based control schemes (\textcolor{orange}{P55076}).

\begin{tcolorbox}[
    colback=gray!8,
    colframe=black,
    width=\linewidth,
    arc=1mm, 
    auto outer arc,
    boxrule=0.8pt,
    boxsep=1pt,
    left=2pt,
    right=2pt,
    top=2pt,
    bottom=2pt
]
\textbf{Finding 1:} {The analysis reveals substantial overlap between user priorities and developer efforts, particularly in areas such as input support, hardware compatibility, tracking, privacy, and immersive features. This alignment indicates a continuous feedback loop, where user experiences and expectations inform technical solutions, and developer innovations in turn shape user perceptions and adoption. Strengthening the interplay between users and developers through transparent communication, responsive design iterations, and a shared focus on emerging challenges can accelerate progress toward more effective and user-aligned VR systems.}
\end{tcolorbox}

\textbf{User Focus.} In addition to the overlapping topics, several user-specific concerns do not appear in the developer discussions. A prominent issue is \textbf{LGBTQ+ representation} (1.9\%). Some users reported experiencing LGBTQ+-related discrimination, criticizing the VR community management as unfair (\textcolor{blue}{R138371}, \textcolor{blue}{R183044302}), while others expressed discomfort with what they perceived as excessive LGBTQ+ promotion in games (\textcolor{blue}{R135166}). Discussions also address \textbf{children, students, and females} (4.4\%). Users tend to highlight the lack of VR games tailored to children (\textcolor{blue}{R87842}) and recognize VR’s potential to enhance realistic learning experiences for students (\textcolor{blue}{R57457}). This aligns with developer efforts to explore VR applications in educational training (\textcolor{orange}{P2478}). In addition, users suggest that increasing female character representation and expanding avatar customization could make VR gaming more appealing to female players (\textcolor{blue}{R194495}, \textcolor{blue}{R248678}).

Another key concern is \textbf{language support} (1.7\%). Users criticize the limited language options in VR games, e.g., the absence of Dutch language support (\textcolor{blue}{R234861}). In addition, pricing issues are frequently debated, with users evaluating game quality relative to cost (\textcolor{blue}{R140819314}). Finally, discussions focus on \textbf{game-specific topics} (4.2\%). For example, in golf-related games, users request more diverse course designs (\textcolor{blue}{R187950}), and in dynamic action games, users point out that movement restrictions, e.g., prolonged standing requirements, create accessibility barriers for players with mobility limitations. To address such issues and improve accessibility, users propose alternative control methods, e.g., alternative button inputs or specific actions (\textcolor{blue}{R213644}).

\begin{tcolorbox}[
    colback=gray!8,
    colframe=black,
    width=\linewidth,
    arc=1mm, 
    auto outer arc,
    boxrule=0.8pt,
    boxsep=1pt,
    left=2pt,
    right=2pt,
    top=2pt,
    bottom=2pt
]
\textbf{Finding 2:} {User discussions highlight a growing demand for greater inclusivity and accessibility in VR, spanning fair LGBTQ+ representation, child-friendly content, expanded language support, and accommodations for players with diverse physical abilities. Addressing these concerns is essential for fostering a more equitable and engaging VR environment. Future research and development should prioritize inclusive design, diverse content, and accessible interaction mechanisms to broaden participation and ensure that VR experiences are welcoming to all user groups.}
\end{tcolorbox}

\textbf{Developer Focus.} Developer forums highlight several technical challenges encountered during the development process. A key area of discussion is \textbf{development tools} (4.0\%), as efficient tools significantly enhance productivity (\textcolor{orange}{P1073653}, \textcolor{orange}{P650628}). \textbf{UE development} (2.8\%) is another prominent topic, with developers addressing cross-platform development, environment setup, and interface design. Issues with tools like Unreal Engine and Unity, particularly compatibility problems arising from version updates, are frequently reported (\textcolor{orange}{P10239}). \textbf{Device compatibility} (2.1\%) remains a critical concern, with developers proposing solutions such as delaying head pose updates until the camera latches the reference frame to address rendering challenges (\textcolor{orange}{P316431}). \textbf{Server management} (2.8\%) is also widely discussed, as larger projects require robust server infrastructure to handle increasing resource demands (\textcolor{orange}{P332852}).

Developers also face \textbf{layout challenges} (2.8\%), e.g., design constraints related to Hydra (\textcolor{orange}{P70745}), and \textbf{support challenges} (2.3\%), including plugin compatibility issues with hardware, as demonstrated in inquiries about DK2 support (\textcolor{orange}{P235055}). Another significant topic is \textbf{cybersickness} (2.5\%), where developers analyze the factors that contribute to motion discomfort and propose design solutions. Some discussions reference Gavgani’s \cite{gavgani2017profiling} work on VR-induced cybersickness (\textcolor{orange}{P585178}), and these technical discussions align with user concerns regarding \textbf{comfort issues}, e.g., reports of discomfort during VR use (\textcolor{blue}{R722882}).

\begin{tcolorbox}[
    colback=gray!8,
    colframe=black,
    width=\linewidth,
    arc=1mm, 
    auto outer arc,
    boxrule=0.8pt,
    boxsep=1pt,
    left=2pt,
    right=2pt,
    top=2pt,
    bottom=2pt
]
\textbf{Finding 3:} {Developer discussions reveal a strong focus on overcoming technical barriers in VR, including the adoption of efficient development tools, ensuring device and platform compatibility, managing server resources, and addressing layout and plugin support challenges. Notably, developers are actively engaged in mitigating cybersickness and enhancing user comfort, aligning technical solutions with end-user needs. These insights highlight the necessity for continued innovation in toolchains, robust compatibility frameworks, and user-centered design strategies. Future research and development should prioritize integrated approaches that streamline development workflows, facilitate cross-platform support, and proactively address factors affecting user comfort and health, ultimately promoting the creation of seamless and enjoyable VR experiences.}
\end{tcolorbox}

\subsection{(RQ2) How topics evolve over time from different perspectives ?}

\subsubsection{Motivation}
Discussions in the VR domain exhibit unique temporal dynamics because user concerns and developer challenges evolve with emerging technological advancements. For example, user feedback regarding hardware devices may shift with iterations in HMD technology, and developer discussions on technical obstacles may decline as solutions mature. 
By addressing RQ2, our goal is to investigate how user needs and topics evolve, resulting in a continuously expanding VR community.

\subsubsection{Approach}
We employ absolute and relative impact metrics in different time periods, which is inspired by Han et al. \cite{han2020programmers} and Lee et al. \cite{uddin2021empirical} to examine dynamic changes in topics within specific fields (e.g., IoT).

\begin{figure*}
	\centering
	\includegraphics[width=\textwidth, height=0.7\textheight]{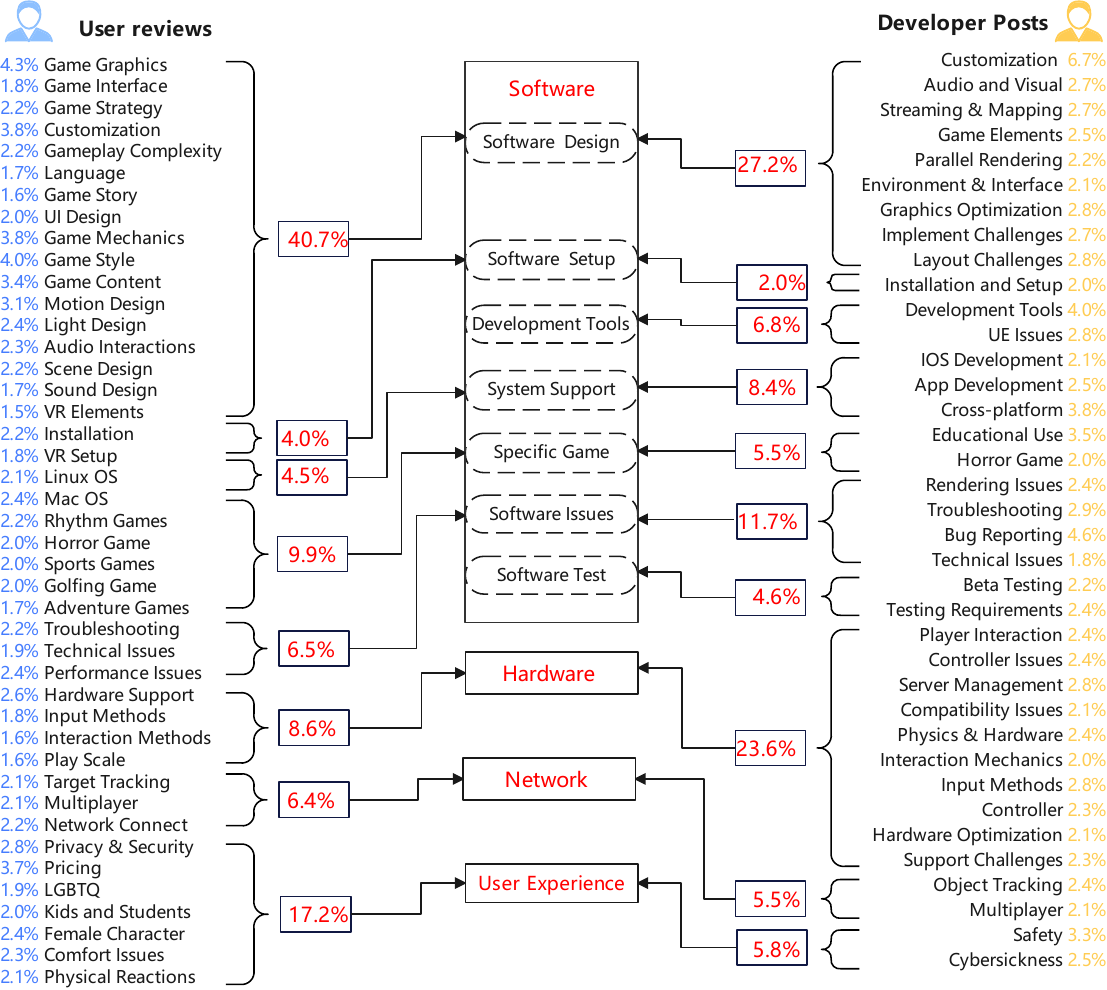}
	\caption{VR Topics with Categories and Subcategories}
	\label{RQ1}
\end{figure*}

We adopt the widely recognized open-coding procedure~\cite{creswell2016qualitative} for the manual inspection and categorization of topics into higher-level categories, as suggested in~\cite{uddin2021empirical}. For example, topics such as \textit{game graphics} and \textit{UI design}, both related to game design, are categorized under the broader "software" category. This iterative process preserves the distinct characteristics of the dataset while consolidating similar topics into more representative and comprehensive categories. The classification is conducted independently by two authors to ensure objectivity and consistency. The iteration continues until consensus is reached among the evaluators and no further optimization can be made. As a result, all topics are organized into four major categories, as shown in Figure~\ref{RQ1}. 


\textbf{Calculation of Absolute Impact.} First, we apply topic popularity metrics to compute the popularity of a topic $z_t$ in corpus $c_j$ for a post or review $d_i$, where $i$ represents any topic in corpus $c_j$. Formally, the popularity of each topic is defined as follows.

\begin{equation}
\begin{split}
\text{popularity}(z_t, c_j) &= \frac{|d_i|}{|c_j|}: \\
\text{dominant}(d_i) &= z_t, \; 1 \leq i \leq c_j, \; 1 \leq j \leq T
\end{split}
\label{Equation_Start}
\end{equation}

We apply LDA to corpus $c_j$ to obtain a set of $T$ topics ($z_1, \ldots, z_t$). We express the probability for a specific topic $z_t$ in a post or review $d_i$ as $\theta(d_i, z_t)$ to define the absolute impact metric of a topic $z_t$ in a month $m$ as follows:

\begin{equation}
\text{impact}_{\text{absolute}}(z_t; m) = \sum_{d_i=1}^{D(m)} \theta(d_i; z_t)
\end{equation}
where $D(m)$ denotes the total number of reviews or posts in month $m$, and $\theta(d_i; z_t)$ represents the probability that the post or review $d_i$ belongs to topic $z_t$.

We further extend this definition to calculate the absolute impact of a category $C$ as follows:
\begin{equation}
\text{impact}_{\text{absolute}}(C; m) = \sum_{z_t \in C} \text{impact}_{\text{absolute}}(z_t; m)
\end{equation}

The category $C$  belongs to four major category of VR topic i.e., Hardware, Software, Network and User Experience.

\textbf{Calculation of Relative Impact.} The relative impact measures the proportion of a topic \(z_t\) relative to all discussions during a specific time. It is defined as:
\begin{equation}
\text{impact}_{\text{relative}}(z_t; m) = \frac{1}{|D(t)|} \sum_{d_i=1}^{D(m)} \theta(d_i; z_t)
\label{Relative}
\end{equation}
where \(|D(t)|\) represents the total number of posts in a mounth \(m\). The \(\theta\) shows the probability of a particular topic \(z_t\) for a post or reveiw \(d_i\). 
The relative impact metric estimates the proportion of posts for a specific topic \(z_t\) relative to all posts or reviews in a particular month \(m\). 

Similarly, the relative impact of a category \(C\) can be calculated as:
\begin{equation}
\text{impact}_{\text{relative}}(C; m) = \sum_{z_t \in C} \text{impact}_{\text{relative}}(z_t; t)
\label{Equation_End}
\end{equation}
where $C$ is the set of posts related to one of the four major categories of posts or reviews.

\subsubsection{Results}

\begin{figure*}[t]
    \centering
    \subfigure[User] {
        \includegraphics[width=.48\textwidth]{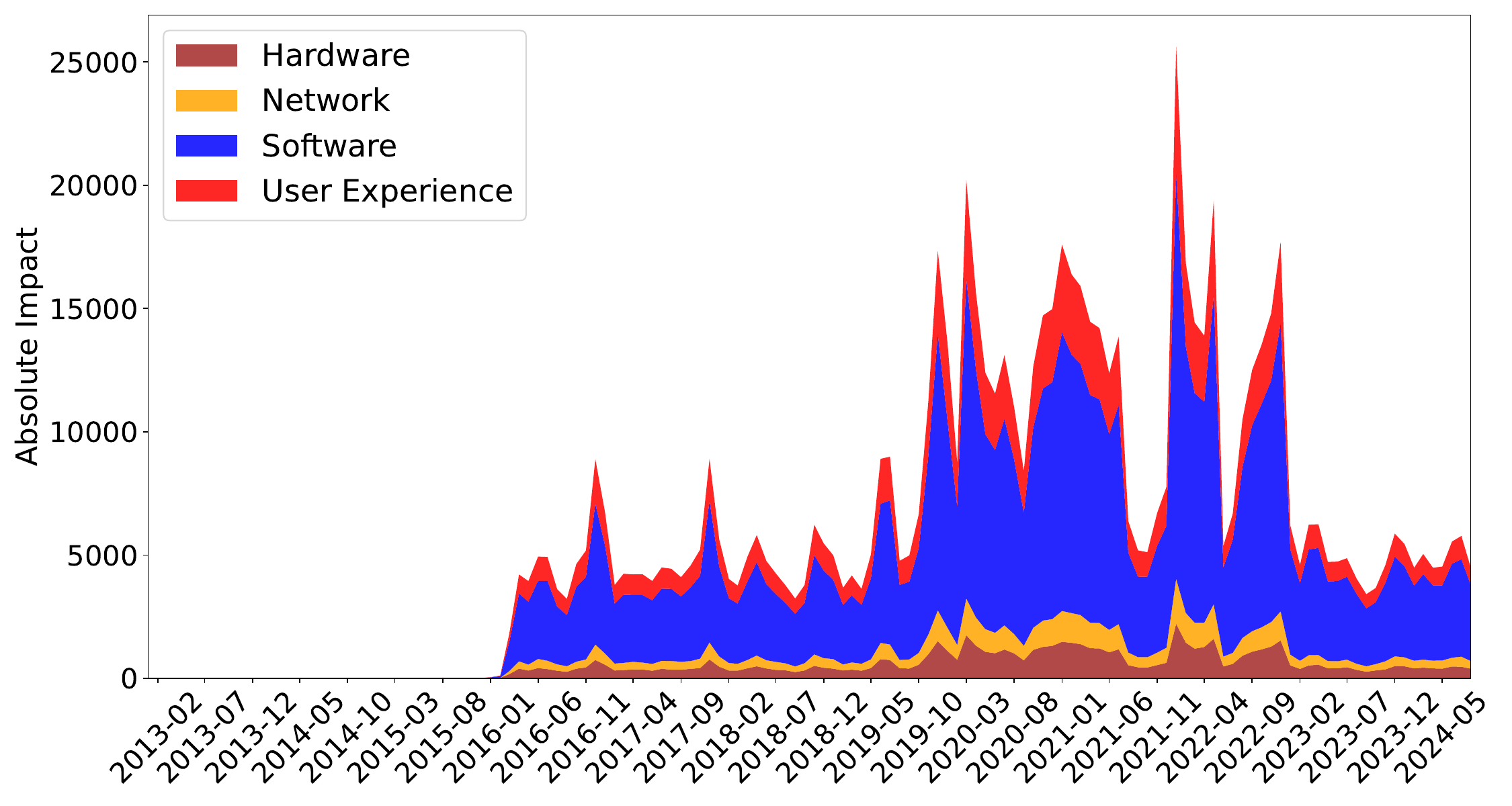}
        \label{fig:Auser}  
    }
    \subfigure[Developer] {
        \includegraphics[width=.48\textwidth]{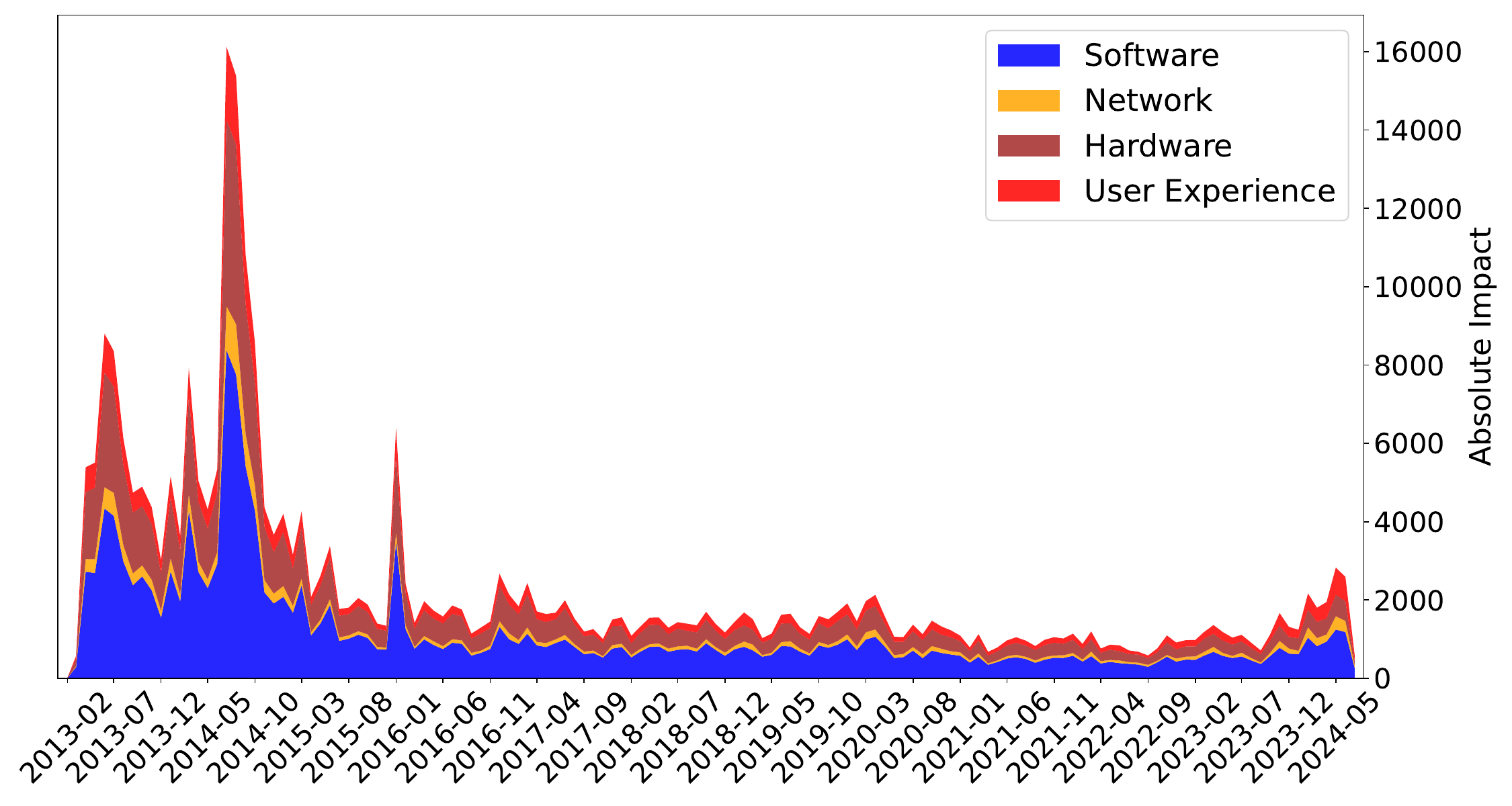}
        \label{fig:Adeveloper}  
    }
\caption{Temporal trends in absolute impact scores of topic popularity for $D_{user}$ and $D_{developer}$ (2013–2024). The X-axis is divided by month, with labels shown at five-month intervals, and the Y-axis represents the absolute impact scores of topics over time.}

    \label{fig:absolute_impact}  
\end{figure*}

\begin{figure*}[t]
    \centering
    \subfigure[User] {
        \includegraphics[width=.48\textwidth]{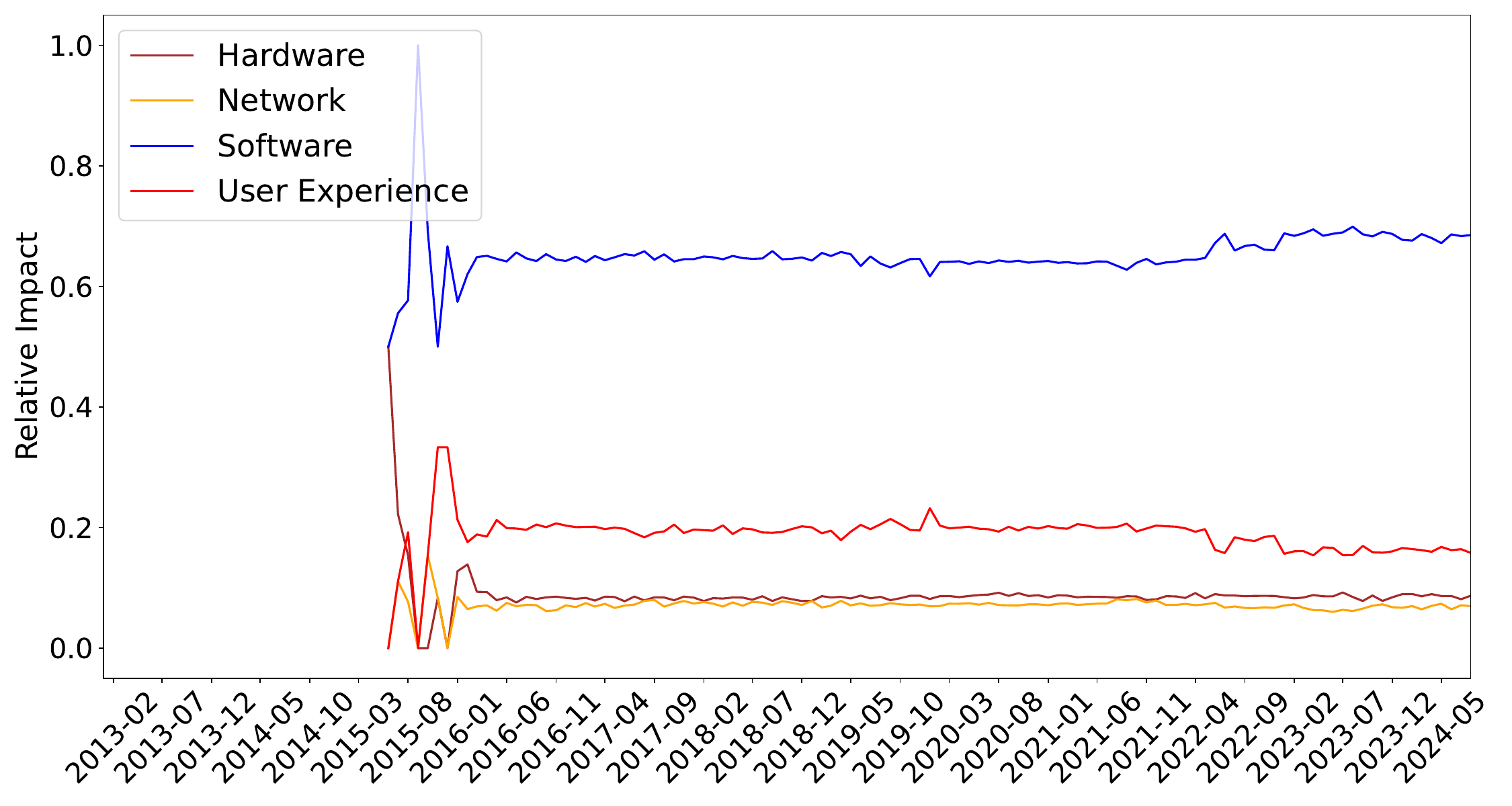}
        \label{fig:Auser}  
    }
    \subfigure[Developer] {
        \includegraphics[width=.48\textwidth]{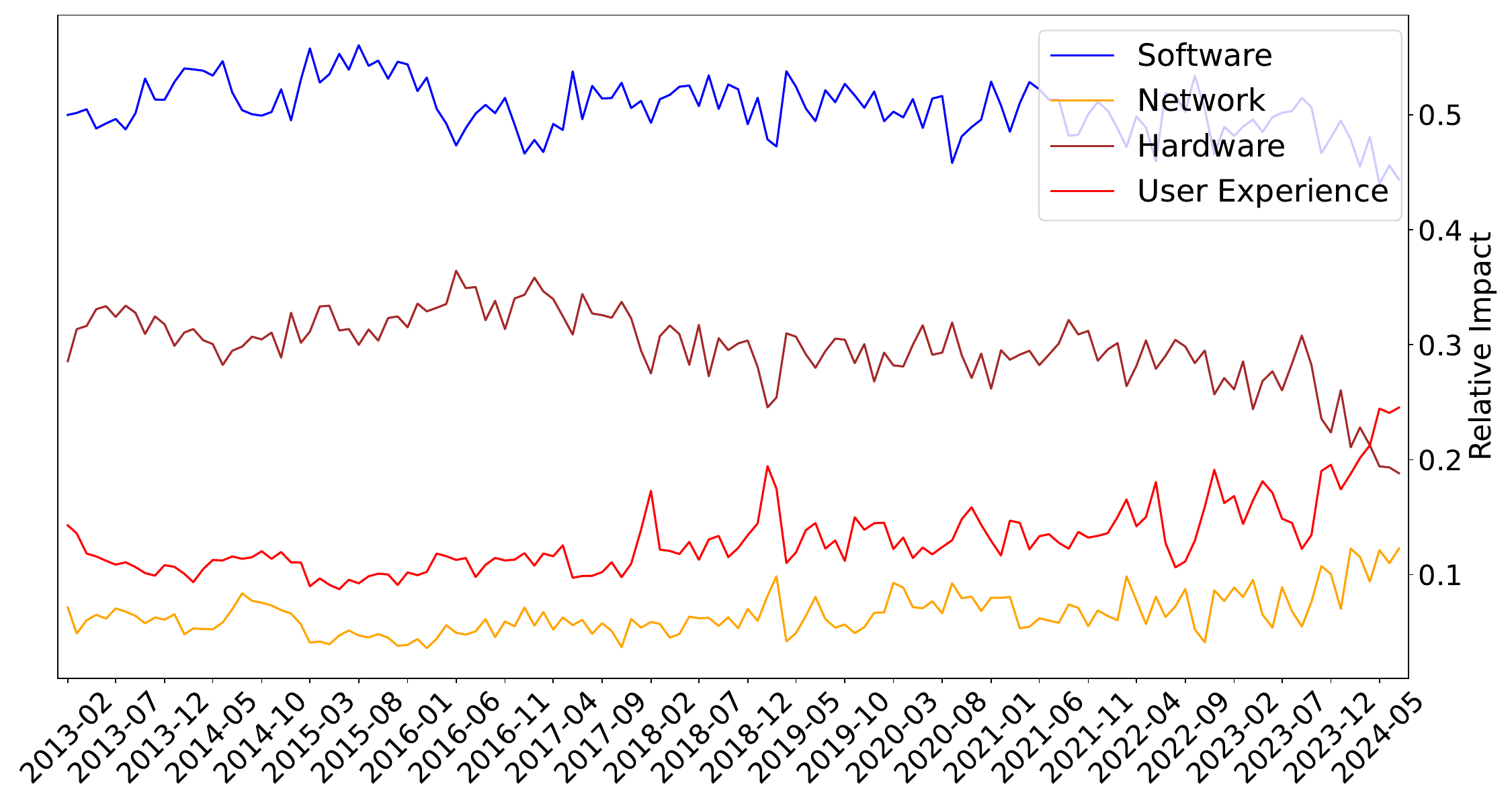}
        \label{fig:Adeveloper}  
    }
\caption{Temporal trends in relative impact scores of topic popularity for $D_{user}$ and $D_{developer}$ (2013–2024). The X-axis is divided by month, with labels shown at five-month intervals, and the Y-axis represents the relative impact scores, indicating the proportion of attention each topic received over time.}

    \label{fig:relative_impact}  
\end{figure*}

The impact of VR topics and categories was quantified using Eqs.(\ref{Equation_Start})–(\ref{Equation_End}), enabling dynamic analysis across four categories, i.e., software, hardware, network, and user experience, from January 2013 to July 2024, as shown in Figure\ref{fig:absolute_impact} and Figure~\ref{fig:relative_impact}.
In the following descriptions, we examine trends in absolute and relative impacts for these categories in both user and developer datasets. For absolute impact, we identify key inflection points and major transitions within each category. For relative impact, we highlight interaction patterns and fluctuations among categories, focusing on periods of overlap, divergence, and sudden shifts.

\textbf{Topic Absolute Impact from User.} Figure~\ref{fig:absolute_impact} (a) illustrates the temporal variation of absolute impact across the four major categories from 2015 to 2024. Software-related topics consistently dominate user discussions, maintaining the highest absolute impact throughout the period. While the overall pattern remains stable, a pronounced surge in absolute impact is observed from late 2019 to early 2022, followed by a return to more stable levels.

\textbf{Software.}
In \textbf{August 2016}, users expressed concerns about visual clarity, with blurry graphics negatively affecting immersion (\textcolor{blue}{R502859}). By \textbf{December 2017}, customization emerged as a central topic, with players requesting greater control over movement mechanics (\textcolor{blue}{R674926}). In \textbf{December 2019}, users shifted focus to game design, calling for more intelligent NPCs and an increased number of levels to enhance engagement (\textcolor{blue}{R735575}). In \textbf{March 2020}, the emphasis turned to content depth, as users demanded expanded gameplay functionalities (\textcolor{blue}{R387783}). By \textbf{January 2021}, users reported frustrations with the complexity of VR setup processes, citing them as barriers to accessibility (\textcolor{blue}{R385168}). Customizable shortcuts were emphasized again in \textbf{January 2022}, with users highlighting their potential to improve usability and efficiency (\textcolor{blue}{R99936}). Most recently, in \textbf{December 2023}, software update failures became a major concern, as players reported that games became unplayable due to malfunctioning updates (\textcolor{blue}{R152968406}).


\textbf{Hardware.} 
In \textbf{August 2016}, controller compatibility issues were widely reported, particularly mismatches between hardware and VR software environments (\textcolor{blue}{R48028}). By \textbf{December 2017}, users began requesting broader support for diverse input devices such as steering wheels and flight controllers (\textcolor{blue}{R681056}). In \textbf{December 2019}, platform compatibility concerns emerged, with users criticizing Oculus for lacking cross-platform support, which limited accessibility (\textcolor{blue}{R76071}). In \textbf{March 2020}, high system requirements frustrated users, who found VR inaccessible due to hardware demands (\textcolor{blue}{R557044}). By \textbf{January 2021}, ergonomic design became a focal point, with complaints about hand controllers like the Vive wand being misaligned with natural hand positions (\textcolor{blue}{R386869}). In \textbf{January 2022}, users emphasized limitations in storage capacity on standalone headsets, which affected performance and load times (\textcolor{blue}{R752039}). Most recently, in \textbf{December 2023}, HMD–software compatibility issues resurfaced, as users reported failures when using certain headsets with specific applications (\textcolor{blue}{R152859220}).

\textbf{Network.} Users initially reported head tracking deficiencies in \textbf{August 2016}, highlighting precision and responsiveness issues that significantly impacted immersion (\textcolor{blue}{R506471}). By \textbf{December 2017}, discussions shifted towards multiplayer modes and cross-brand controller connectivity problems (\textcolor{blue}{R12218}). Network instability, including frequent latency spikes and disconnections, became a critical topic by \textbf{December 2019} (\textcolor{blue}{R05846}). User concerns in \textbf{March 2020} emphasized sluggish target-tracking responsiveness, noting that delayed inputs severely disrupted immersive experiences (\textcolor{blue}{R562457}). Bluetooth convenience and reliability emerged as key discussion points by \textbf{January 2021} (\textcolor{blue}{R237800}). By \textbf{January 2022}, users increasingly reported issues related to hand tracking accuracy, expressing frustration over frequent recognition errors (\textcolor{blue}{R277222}). Most recently, in \textbf{December 2023}, users highlighted limitations in camera tracking compatibility with HOTAS devices, affecting gameplay in specialized VR applications (\textcolor{blue}{R54298220}).

\textbf{User Experience.} Early VR users frequently discussed physiological discomfort, notably VR-induced nausea and dizziness in \textbf{August 2016} (\textcolor{blue}{R320532}). Economic considerations regarding the cost-effectiveness of VR hardware and software became prevalent in \textbf{December 2017}, indicating users’ concerns about long-term investment value (\textcolor{blue}{R505014}). By \textbf{December 2019}, the increasing participation of younger users led to significant discussions about community management, especially regarding behavior moderation to ensure immersive experiences for all (\textcolor{blue}{R324096}). By \textbf{March 2020}, the inclusivity of environmental settings, particularly for child players, was critically reviewed, with suggestions for more adaptive design (\textcolor{blue}{R270975}). Social inclusivity became increasingly prominent in \textbf{January 2021}, with specific calls to address discriminatory behavior towards LGBTQ+ users in social VR platforms (\textcolor{blue}{R140921}). Data privacy became a dominant concern by \textbf{January 2022}, highlighting debates about ethical data handling and user privacy protections in VR platforms like Facebook (\textcolor{blue}{R142345}). In the latest discussions of \textbf{December 2023}, users advocated for improved physics engines to enhance immersive realism through more precise physical feedback (\textcolor{blue}{R154626566}).

\begin{tcolorbox}[
    colback=gray!8,
    colframe=black,
    width=\linewidth,
    arc=1mm, 
    auto outer arc,
    boxrule=0.8pt,
    boxsep=1pt,
    left=2pt,
    right=2pt,
    top=2pt,
    bottom=2pt
]               
\textbf{Finding 4:} {This study shows a clear shift in user concerns from basic functionality (“Can it work?”) to quality and experience (“Does it work well?”), reflecting higher expectations as VR technology matures. User discussions have moved from feasibility and usability to stability, smooth interaction, social inclusivity, and immersion. For future research and development, it is essential to go beyond basic functionality and create VR systems that are reliable, user-friendly, and inclusive. Developers and researchers should focus on improving software reliability, ergonomic hardware design, and tailored user experiences to support broader adoption and long-term satisfaction.}
\end{tcolorbox}


\textbf{Topic Absolute Impact from Developer.} Figure~\ref{fig:absolute_impact} (b) presents the absolute impact trends across four categories from 2013 to 2024. Software topics overwhelmingly dominate developer discussions, especially during the early years, with a pronounced peak around 2014–2015. Following this surge, the absolute impact across all categories declines sharply and stabilizes at a much lower level, with only minor fluctuations observed in recent years.

\textbf{Software.} Software discussions showed evolving priorities, beginning in \textbf{June 2013} with compatibility concerns, exemplified by Deskope supporting Rift with Windows (\textcolor{orange}{P105980}). By \textbf{August 2014}, attention shifted towards video playback, particularly DK2's Video Player functionality (\textcolor{orange}{P193258}). Rendering issues, notably SpaceEngine in direct mode, became central in \textbf{December 2014} (\textcolor{orange}{P284695}). Audio configuration discussions emerged in \textbf{January 2016}, emphasizing stereo versus surround sound preferences (\textcolor{orange}{P342270}). By \textbf{May 2024}, developers tackled compatibility issues like Quest 3 and Unreal Engine 5.3.2 passthrough displays showing black screens (\textcolor{orange}{P1193769}). 

\textbf{Hardware.} 
Initially, in \textbf{June 2013}, high-resolution devices were prioritized (\textcolor{orange}{P80843}). Compatibility discussions, particularly regarding the Digital Combat Simulator with DK2, gained prominence in \textbf{August 2014} (\textcolor{orange}{P181179}). Attention moved to accessory improvements such as Rift lens replacements in \textbf{December 2014} (\textcolor{orange}{P284366}). GPU optimization discussions were significant by \textbf{January 2016} (\textcolor{orange}{P377721}). Recent hardware discussions in \textbf{May 2024} included multimodal interaction challenges with Oculus controllers (\textcolor{orange}{P1199093}). 
\textbf{Network.}
Initial conversations in \textbf{June 2013} explored Wi-Fi-based tracking feasibility (\textcolor{orange}{P75710}). By \textbf{August 2014}, developers discussed controller connectivity and latency issues (\textcolor{orange}{P217100}). Optimization of DK2 connectivity became prominent by \textbf{December 2014} (\textcolor{orange}{P332607}). Tracking accuracy improvements, such as increased LED use in Oculus DK2, dominated in \textbf{January 2016} (\textcolor{orange}{P324142}). Renewed attention to motion capture and body tracking characterized discussions in \textbf{May 2024} (\textcolor{orange}{P1196371}). 

\textbf{User Experience.} UX discussions evolved from basic comfort to advanced personalization. Cybersickness mitigation, such as Tai Chi applications, was a focus in \textbf{June 2013} (\textcolor{orange}{P61346}). By \textbf{August 2014}, attention turned to screen door effects and visual discomfort (\textcolor{orange}{P226422}). Interpupillary distance adjustments for motion sickness were central in \textbf{December 2014} (\textcolor{orange}{P330891}). Rotational comfort strategies gained prominence in \textbf{January 2016} (\textcolor{orange}{P346572}). By \textbf{May 2024}, discussions shifted towards user-customized VR solutions (\textcolor{orange}{P1194513}). 

\begin{tcolorbox}[
colback=gray!8,
colframe=black,
width=\linewidth,
arc=1mm,
auto outer arc,
boxrule=0.8pt,
boxsep=1pt,
left=2pt,
right=2pt,
top=2pt,
bottom=2pt
]\textbf{Finding 5:} {VR developer discussions reflect a clear evolution from basic functional concerns ("Will it work?") to sophisticated user-centric enhancements ("How can we make it awesome?"). Trends indicate a significant shift towards software integration, hardware optimization, robust networking, and personalized UX solutions, aligning with technological advances and user expectations. Future research and development should prioritize integrated system approaches, improved connectivity, and highly customized user experiences to foster sustained adoption and satisfaction.}
\end{tcolorbox}

\textbf{ Topic Relative Impact from User.} Figure \ref{fig:relative_impact}(a) presents the relative changes in topic popularity, illustrating the distribution and temporal evolution of each topic within these categories. Since \textbf{early 2015}, user interests on VR platforms have experienced rapid differentiation, eventually stabilizing into a structure dominated by \textbf{Software} topics, with \textbf{User Experience} following closely behind. The \textbf{Software} category has consistently maintained approximately \textbf{60\%} of the relative impact, serving as the core of community discussions, while \textbf{User Experience} has remained stable at around \textbf{20\%--25\%}. In contrast, both \textbf{Hardware} and \textbf{Network} topics have consistently accounted for less than \textbf{10\%} each, indicating their marginal status within the overall discourse.

During \textbf{April 2017}, \textbf{October 2017}, \textbf{September 2018}, and \textbf{August 2021}, the topics of \textbf{Hardware} and \textbf{Network} repeatedly exhibited convergence and simultaneous surges in user attention. This phenomenon reflects users' dual focus on VR device performance upgrades and the optimization of online experiences. For example, users anticipated new gameplay enabled by enhanced hardware capabilities (e.g., GTX 1060 support, \textcolor{blue}{R465148}) and simultaneously showed increased sensitivity and controversy regarding network experiences following platform updates (\textcolor{blue}{R54447}). Over time, users placed greater emphasis on system-level synergy among content depth, device capability, and network services, as seen in discussions on desktop mapping and highly immersive interaction (\textcolor{blue}{R332041}, \textcolor{blue}{R615691}). Users frequently called for holistic, end-to-end optimization in response to the tension between hardware advancements and network service limitations. By \textbf{August 2021}, a diversified experiential system characterized by community activity, content innovation, and platform collaboration had become the standard for high-quality VR experiences. The integration of \textbf{Hardware}, \textbf{Network}, and community engagement continued to drive the evolution of the VR ecosystem, as illustrated by discussions on MOD experiences (\textcolor{blue}{R621037}) and the sense of social belonging (\textcolor{blue}{R616620}).

In \textbf{September 2019}, \textbf{March 2022}, and \textbf{August 2022}, the topics of \textbf{User Experience} and \textbf{Software} exhibited alternating rises and differentiation. When \textbf{User Experience} increased, users tended to raise higher expectations regarding onboarding, operational details, and fairness (e.g., lack of tutorials, \textcolor{blue}{R135539}), while the \textbf{Software} topic focused on emotional resonance, aesthetics, and enjoyment (e.g., \textcolor{blue}{R736720}). With improvements in platform functionality and compatibility, the \textbf{Software} topic received positive feedback for technological advancements (such as interface design, visual quality, and functional innovation, \textcolor{blue}{R120065517}, \textcolor{blue}{R300787}); however, when content depth and diversity were lacking, the \textbf{User Experience} topic declined due to issues like short storylines or repetitive tasks (\textcolor{blue}{R121041547}, \textcolor{blue}{R268013}).

Entering \textbf{2024}, the focal points of \textbf{User Experience} and \textbf{Software} topics have demonstrated both deep integration and further differentiation. On the \textbf{User Experience} front, extremely positive feedback and diverse demands have become mainstream (e.g., ``best VR game,'' \textcolor{blue}{R167402269}; ``peaceful mode,'' \textcolor{blue}{R166799844}), alongside a growing user demand for emotion regulation and extreme gameplay experiences. Meanwhile, the \textbf{Software} topic features a remarkably high degree of interaction freedom and emotional release (e.g., ``sadistic rage,'' \textcolor{blue}{R168182744}), as well as reflections on moral and social issues (e.g., \textcolor{blue}{R168235987}).

\begin{tcolorbox}[
colback=gray!8,
colframe=black,
width=\linewidth,
arc=1mm,
auto outer arc,
boxrule=0.8pt,
boxsep=1pt,
left=2pt,
right=2pt,
top=2pt,
bottom=2pt
]\textbf{Finding 6:} {The long-term evolution of user interests on VR platforms reveals a persistent emphasis on \textbf{software functionality and innovation}, alongside increasing attention to \textbf{user experience}, while \textbf{hardware} and \textbf{network} periodically become prominent during ecosystem upgrades. As user expectations expand from isolated features to integrated “content-hardware-network-community” experiences and emotional value, achieving high-quality VR will require balancing core usability, content richness, emotional engagement, and systemic synergy. }
\end{tcolorbox}

\textbf{Topic Relative Impact from \textcolor{orange}{Developer}.} Figure~\ref{fig:relative_impact} (b) illustrates the distribution and temporal evolution of topic popularity across categories. From \textbf{2013} to \textbf{2024}, developer forum discussions exhibited pronounced cyclical trends and distinct evolutionary phases. \textbf{Software} consistently dominated the discourse, accounting for \textbf{45\%--55\%} of attention. \textbf{Hardware} initially alternated with software as the primary focus, but after \textbf{2015} its prominence stabilized at \textbf{25\%--35\%}. \textbf{User Experience} demonstrated continuous growth, rising to approximately \textbf{20\%} by \textbf{2024} and becoming the second most prominent topic. In contrast, \textbf{Network} maintained a relatively low profile, generally below \textbf{10\%}, except for occasional surges during major platform events.

Between \textbf{April 2013} and \textbf{December 2013}, \textbf{Software} topics increased markedly, with developers focusing on the evaluation and implementation of emerging platform tools (e.g., Vorpx VR middleware), API integration, and content ecosystems (e.g., Mate), which stimulated vibrant community discussion (\textcolor{orange}{P45260}, \textcolor{orange}{P94111}, \textcolor{orange}{P132473}). In contrast, interest in \textbf{Hardware} declined, centering on the performance, peripheral integration, and compatibility of headsets such as the Rift (\textcolor{orange}{P50428}, \textcolor{orange}{P110594}, \textcolor{orange}{P128752}), indicative of a stabilization in hardware development. \textbf{User Experience} topics also decreased, with discussions on gesture tracking, embodied interaction, and environmental adaptation (e.g., Leap Motion, \textcolor{orange}{P109799}, \textcolor{orange}{P34733}, \textcolor{orange}{P27161}) becoming more diffuse as interaction paradigms matured. \textbf{Network} remained the least discussed, with only brief surges during device adaptation and low-level debugging (e.g., Rift network configuration across different computers and resolutions, \textcolor{orange}{P35154}, \textcolor{orange}{P40218}, \textcolor{orange}{P114457}), further subsiding as technical challenges were addressed.

From \textbf{June 2014} to \textbf{December 2015}, the release of next-generation hardware platforms (such as Oculus CV1) and advancements in core parameters like FoV and resolution became the main drivers of heightened developer interest. The prominence of \textbf{Hardware} topics increased sharply, with community discussions focusing on device compatibility, peripheral integration, and proof-of-concept validation (\textcolor{orange}{P160496}, \textcolor{orange}{P186772}, \textcolor{orange}{P193366}). Meanwhile, \textbf{Software} topics gradually stabilized, as developers concentrated on content innovation and ecosystem compatibility (e.g., new versions of the Mate platform, \textcolor{orange}{P255485}, \textcolor{orange}{P227487}, \textcolor{orange}{P145477}); the toolchain matured, and the pace of innovation slowed slightly. \textbf{User Experience} discussions centered on flagship titles (such as Oculus and Half-Life 2), with a focus on immersive content and product feedback, though overall attention declined (\textcolor{orange}{P197393}, \textcolor{orange}{P332755}, \textcolor{orange}{P320324}), marking a community shift towards hardware-centric concerns. \textbf{Network} topics primarily addressed community management, authentication, and account system development, with relatively limited engagement (\textcolor{orange}{P224722}, \textcolor{orange}{P242079}, \textcolor{orange}{P148747}).

In \textbf{October 2018}, ecosystem synergy and diversified content innovation emerged as new focal points within the community. The prominence of \textbf{Network} and \textbf{User Experience} topics increased significantly, with developers actively discussing multi-location collaboration, content synchronization, and the launch of new services on platforms such as Oculus and SteamVR, thereby advancing cross-platform interoperability (\textcolor{orange}{P731934}, \textcolor{orange}{P676749}, \textcolor{orange}{P696682}). On the \textbf{User Experience} front, content uploading, device compatibility, and community events became central concerns, and there was a clear rise in demands for ecosystem openness and interactive innovation (e.g., Google Drive integration, device list upgrades, \textcolor{orange}{P21382}, \textcolor{orange}{P692004}, \textcolor{orange}{P676755}). In contrast, attention to \textbf{Software} and \textbf{Hardware} topics waned, with community discussion focusing more on routine issues such as APIs, content distribution channels, and device crash resolution (\textcolor{orange}{P21317}, \textcolor{orange}{P21335}, \textcolor{orange}{P21147}, \textcolor{orange}{P21623}, \textcolor{orange}{\textsc{P698394}}).

In \textbf{April 2022}, focus on \textbf{User Experience} topics declined, as developers concentrated on controller comfort, interaction commands, and scenario details for devices like VIVE (e.g., Vive palm pinch, peripheral evaluations, \textcolor{orange}{P47320}, \textcolor{orange}{P47150}), reflecting mature basic interactions and device adaptation. The lack of major breakthroughs led to reduced discussion intensity. By \textbf{November 2022}, \textbf{User Experience} surpassed hardware as the second most discussed topic. \textbf{Hardware} discussions addressed bug fixes, display issues, and compatibility (\textcolor{orange}{P50113}, \textcolor{orange}{P50034}, \textcolor{orange}{P998953}), while \textbf{User Experience} centered on detail optimization, third-party integration, and community Q\&A (e.g., Pimax compatibility, soundscape commands,\textcolor{orange}{P995868}).

From \textbf{September 2022} to \textbf{June 2024}, community discussions developed a dual-core structure centered on \textbf{Software} innovation and \textbf{User Experience}. \textbf{Software} topics focused on custom development and multi-platform integration for emerging platforms such as OpenXR, Unity, and SteamVR APIs ( \textcolor{orange}{P170249942}, \textcolor{orange}{P170126759}), with engine features and content distribution as mainstream concerns. \textbf{User Experience} continued to rise, encompassing multi-scenario adaptation, interaction optimization, and community support for devices like Meta Quest and Pimax (\textcolor{orange}{P999795}, \textcolor{orange}{P994961}, \textcolor{orange}{P995868}). \textbf{Hardware} topics exhibited periodic peaks, mainly related to the release of new HMD models, driver compatibility, and module updates (\textcolor{orange}{P169583019}, \textcolor{orange}{P169755139}, \textcolor{orange}{P170004160}). While \textbf{Network} remained less prominent, it showed notable fluctuations during cross-platform synchronization, cloud services, and collaborative node discussions (\textcolor{orange}{P999199}, \textcolor{orange}{P170054170}, \textcolor{orange}{P995949}).

\begin{tcolorbox}[
colback=gray!8,
colframe=black,
width=\linewidth,
arc=1mm,
auto outer arc,
boxrule=0.8pt,
boxsep=1pt,
left=2pt,
right=2pt,
top=2pt,
bottom=2pt
]
\textbf{Finding 7:} VR developers’ focus has shifted from \textbf{hardware} and \textbf{network} adaptation to \textbf{software innovation} and \textbf{user experience}. These have become the main drivers of ecosystem growth, especially with increased attention to \textbf{multi-platform integration} and \textbf{personalized experiences}. While \textbf{hardware} and \textbf{network} receive less ongoing focus, they regain prominence with major device launches. Sustained community vitality relies on \textbf{ecosystem openness}, \textbf{content innovation}, \textbf{user feedback}, and \textbf{multi-platform connectivity}.
\end{tcolorbox}

\section{Implication}
\label{implication}
This section 
is organized into two parts: the first highlights methodological innovations and their broad applicability, while the second offers actionable recommendations for developers and researchers to address key gaps identified within the evolving VR ecosystem.

\subsection{Methodological Implications: Scalable and Interpretable Text Analytics}

The results verify that the proposed HTModel is an effective and scalable approach for large-scale textual analysis in software engineering. 
First, HTModel achieves higher label precision, greater interpretability, and lower annotation costs than LLM-only pipelines by decoupling clustering from LLM-based semantic refinement. Second, its multidimensional evaluation framework, integrating coherence, coverage, diversity, label accuracy, and label usefulness, ensures robust topic extraction and strong domain adaptability across heterogeneous real-world datasets. Third, HTModel can be extended with neural, hierarchical, or domain-adaptive models to support multilingual and cross-platform contexts, effectively capturing the evolving complexity of software ecosystems.

\subsection{Practical Implications: Enhancing User Experience through Multi-Perspective Insights}

Our empirical findings highlight critical areas where targeted efforts from both developers and researchers can significantly improve the VR ecosystem, enhancing user satisfaction, inclusivity, and long-term platform sustainability. Integrating user and developer perspectives uniquely reveals previously unrecognized challenges, providing actionable guidance extending beyond conventional single-stakeholder studies.

\textbf{Ensure inclusive and adaptive design.} User discussions consistently emphasize the necessity for adaptive interfaces accommodating diverse physical and cognitive abilities. Developers should:
\begin{itemize}
    \item Implement adaptive interfaces that support varying accessibility requirements, reducing barriers for users with physical or cognitive impairments. \textit{For example, in VR fitness games, offer configurable control schemes and gesture sensitivity settings so that both wheelchair users and standing players can fully participate.}
    \item Enable extensive avatar personalization options to authentically represent diverse user identities across cultural, gender, and personal expression dimensions. \textit{For example, allow players to customize avatars with cultural attire, gender-neutral features, or assistive devices such as hearing aids or prosthetic limbs.}
    \item Develop environments specifically designed for vulnerable or underrepresented groups such as children, elderly individuals, and persons with disabilities. \textit{For example, design child-friendly VR classrooms with simplified navigation and content filtering for age-appropriate learning.}
\end{itemize}

\textbf{Enhance moderation and safety frameworks.} Persistent user concerns about inadequate moderation necessitate proactive safety measures. Developers are advised to:
\begin{itemize}
    \item Embed intuitive reporting mechanisms directly into VR platforms to facilitate user-driven moderation. \textit{For example, integrate a ``Report'' button into the user interface that can be activated via a quick gesture or voice command during interactions.}
    \item Integrate automated moderation technologies capable of real-time detection and intervention against harassment or inappropriate content. \textit{For example, use AI-based voice chat monitoring to detect hate speech instantly and mute the offending player until reviewed by a moderator.}
    \item Clearly articulate and rigorously enforce community guidelines to sustain respectful, supportive interactions among users. \textit{For example, display a brief but visible code of conduct whenever users join a multiplayer session.}
\end{itemize}

\textbf{Optimize performance equitably.} Performance inequities stemming from hardware disparities create accessibility barriers. Developers should:
\begin{itemize}
    \item Prioritize performance optimization across a diverse spectrum of devices, including mid- to low-tier VR headsets. \textit{For example, include a ``performance mode'' that automatically reduces rendering load while maintaining gameplay mechanics for budget devices.}
    \item Conduct regular assessments of feature impacts on rendering performance, latency, and overall usability. \textit{For example, run automated benchmark tests for every major update to ensure frame rates remain stable across supported hardware.}
    \item Refrain from over-reliance on high-end hardware specifications when developing new functionalities. \textit{For example, ensure that newly introduced visual effects, such as advanced particle simulations, have simplified fallback versions for less powerful systems.}
\end{itemize}

\subsubsection{Implications For Researchers}

\textbf{Examine user--developer alignment longitudinally.} Researchers must investigate the dynamic interactions between user expectations and developer initiatives. Specifically, researchers are encouraged to:
\begin{itemize}
    \item Conduct longitudinal studies incorporating multi-stakeholder analyses to expose subtle divergences that single-perspective approaches might miss. \textit{For example, track how requests for accessibility features from users are addressed (or ignored) across multiple software release cycles.}
    \item Facilitate early detection of user--developer mismatches, thereby informing timely, evidence-driven interventions. \textit{For example, apply topic modeling quarterly to compare developer forum priorities with user reviews, flagging mismatches in content focus.}
    \item Leverage comprehensive datasets to systematically track gaps between inclusive design demands and developer technical priorities. \textit{For example, map keyword clusters from both sides to quantify underrepresented user demands.}
\end{itemize}

\textbf{Explore the role of inclusivity in user retention.} Sustained user engagement is closely linked to inclusivity in design. Researchers are encouraged to:
\begin{itemize}
    \item Investigate how inclusive design elements such as diverse avatars, accessible interactions, and moderated environments affect long-term user engagement and loyalty. \textit{For example, compare retention metrics between platforms that offer inclusive avatar customization and those that do not.}
    \item Utilize mixed-method and longitudinal research designs to uncover causal relationships between inclusive practices and sustained user retention. \textit{For example, combine large-scale user log data with follow-up surveys to understand why users remain active on certain inclusive platforms.}
\end{itemize}

\textbf{Assess intervention efficacy rigorously.} To bridge identified user--developer gaps, we highlight the need for:
\begin{itemize}
    \item Implement controlled empirical evaluations of strategies designed to address gaps, using structured feedback mechanisms, inclusive quality benchmarks, and participatory design methodologies. \textit{For example, run an A/B test comparing standard onboarding with an adaptive onboarding flow for users with different interaction needs.}
    \item Identify and validate best practices for adaptability and inclusivity in VR through systematic, evidence-based testing. \textit{For example, develop a public benchmark dataset of inclusive design implementations and their observed impact on user satisfaction.}
\end{itemize}

Moreover, researchers should explore emerging topics such as the ethical implications of data privacy in immersive environments and the psychological impacts of prolonged VR exposure, areas that surfaced prominently in user discussions but remain underexplored academically. By addressing these topics, future research can contribute substantially to the ethical and sustainable advancement of VR technology.

\section{Threats to Validity}
\label{limitation}


\subsection{Internal validity}
The main internal threat relates to the reliability of the proposed HTModel, i.e., the effectiveness of topic assignment and label consistency. HTModel integrates efficient clustering algorithms with LLM-based semantic understanding, reducing manual annotation compared to traditional clustering models and lowering annotation costs compared to pure LLM-based approaches. HTModel provides high precision, interpretability, and efficiency. 
However, the number of evaluators and the topics that occur less frequently are easily ignored. For example, although a comprehensive evaluation framework is applied, integrating quantitative metrics such as coherence, coverage, and diversity, and expert-based qualitative assessments such as label accuracy and usefulness, the qualitative evaluation involves only two evaluators. 

To address these limitations, future work should consider integrating domain-adaptive or hierarchical models and semi-supervised learning approaches to improve topic boundary definition and adapt to evolving vocabularies.

\subsection{External validity}
The main threat to external validity concerns the generalizability of our findings beyond the current datasets and domain. Our analysis is based on textual data from three major VR gaming platforms and their associated developer forums. While these platforms cover a substantial portion of the VR ecosystem, they do not capture the full range of VR applications, such as those in education, healthcare, or industrial settings. Additionally, discussions from alternative channels, for example social media platforms or private repositories, are not included and may present different topical distributions or community concerns. The study is also limited to English-language data, with more than 100,000 non-English entries excluded during preprocessing. This language restriction may lead to the omission of region-specific issues and culturally distinct perspectives, as highlighted in previous work \cite{lu2024empirical, dong2024points, epp2021empirical}. 

To address these limitations, future research should include a wider range of platforms, application domains, and languages to better assess the generalizability and applicability of the proposed method and findings.

\section{Conclusion}
\label{conclusion}

This decade-spanning multi-stakeholder analysis of the VR ecosystem reveals both convergences and enduring disparities in priorities between users and developers. By examining 944,320 user reviews alongside 389,477 developer forum posts from Meta, SteamVR, and Viveport, the study finds that certain technical concerns—most notably performance optimization and input methods—consistently emerge as shared focal points. At the same time, critical issues surrounding inclusivity, user safety, and community experience remain markedly under-addressed by developers despite persistent and vocal user concern. Temporal trend analysis further exposes an asynchronous evolution of stakeholder interests: over the years, user discourse has increasingly gravitated toward emotional, inclusive, and social dimensions of VR engagement, whereas developer discussions have remained largely anchored in backend technical improvements and platform integration efforts. This misalignment in focus is not merely anecdotal but empirically evident, underscoring a substantive gap between what users value in emerging VR experiences and where developers are investing their effort. 

Beyond these empirical insights, the study makes a broader scholarly contribution through its introduction of a multi-view analytical framework grounded in a novel hybrid topic modeling method (HTModel). By integrating classical topic modeling with a state-of-the-art language model, HTModel achieved high topic-labeling accuracy (exceeding 80\%) while significantly reducing the need for manual annotation, thus demonstrating a cost-effective approach to large-scale text analytics. Notably, compared to fully LLM-based topic modeling pipelines, HTModel reduces computational cost by over 99\% while maintaining high interpretability and scalability. The resulting framework and the curated cross-platform dataset are made available for reuse, providing valuable tools for further research on software ecosystems.

Overall, The findings and methodological advances presented here not only deepen our understanding of user–developer dynamics in VR but also offer a generalizable approach for examining multi-perspective data in other domains. Looking forward, this work opens important avenues for future investigation, such as extending analyses to non-English and region-specific content, incorporating additional stakeholder groups (e.g., platform designers or community moderators), and developing intervention strategies to better synchronize developer priorities with the inclusive, community-oriented expectations of users.

\bibliographystyle{ACM-Reference-Format}
\bibliography{sample-base}


\begin{thebibliography}{50}


\ifx \showCODEN    \undefined \def \showCODEN     #1{\unskip}     \fi
\ifx \showISBNx    \undefined \def \showISBNx     #1{\unskip}     \fi
\ifx \showISBNxiii \undefined \def \showISBNxiii  #1{\unskip}     \fi
\ifx \showISSN     \undefined \def \showISSN      #1{\unskip}     \fi
\ifx \showLCCN     \undefined \def \showLCCN      #1{\unskip}     \fi
\ifx \shownote     \undefined \def \shownote      #1{#1}          \fi
\ifx \showarticletitle \undefined \def \showarticletitle #1{#1}   \fi
\ifx \showURL      \undefined \def \showURL       {\relax}        \fi
\providecommand\bibfield[2]{#2}
\providecommand\bibinfo[2]{#2}
\providecommand\natexlab[1]{#1}
\providecommand\showeprint[2][]{arXiv:#2}

\bibitem[Angelov(2020)]%
        {angelov2020top2vec}
\bibfield{author}{\bibinfo{person}{Dimo Angelov}.} \bibinfo{year}{2020}\natexlab{}.
\newblock \showarticletitle{Top2vec: Distributed representations of topics}.
\newblock \bibinfo{journal}{\emph{arXiv preprint arXiv:2008.09470}} (\bibinfo{year}{2020}).
\newblock


\bibitem[Ashtari et~al\mbox{.}(2023)]%
        {ashtari2023arvr_challenges}
\bibfield{author}{\bibinfo{person}{Amirhossein Ashtari}, \bibinfo{person}{Michael Nebeling}, {and} \bibinfo{person}{Moritz Speicher}.} \bibinfo{year}{2023}\natexlab{}.
\newblock \showarticletitle{An Empirical Study on Current Practices and Challenges of Core AR/VR Developers}.
\newblock \bibinfo{journal}{\emph{ACM CHI Workshop on AR/VR Development}} (\bibinfo{year}{2023}).
\newblock


\bibitem[Azher et~al\mbox{.}(2024)]%
        {azher2024limtopic}
\bibfield{author}{\bibinfo{person}{Ibrahim~Al Azher}, \bibinfo{person}{Venkata Devesh~Reddy Seethi}, \bibinfo{person}{Akhil~Pandey Akella}, {and} \bibinfo{person}{Hamed Alhoori}.} \bibinfo{year}{2024}\natexlab{}.
\newblock \showarticletitle{Limtopic: Llm-based topic modeling and text summarization for analyzing scientific articles limitations}. In \bibinfo{booktitle}{\emph{Proceedings of the 24th ACM/IEEE Joint Conference on Digital Libraries}}. \bibinfo{pages}{1--12}.
\newblock


\bibitem[Barua et~al\mbox{.}(2014)]%
        {barua2014stackoverflow}
\bibfield{author}{\bibinfo{person}{Anton Barua}, \bibinfo{person}{Stephen~W Thomas}, {and} \bibinfo{person}{Ahmed~E Hassan}.} \bibinfo{year}{2014}\natexlab{}.
\newblock \showarticletitle{What are developers talking about? An analysis of topics and trends in Stack Overflow}.
\newblock \bibinfo{journal}{\emph{Empirical Software Engineering}} \bibinfo{volume}{19}, \bibinfo{number}{3} (\bibinfo{year}{2014}), \bibinfo{pages}{619--654}.
\newblock


\bibitem[Blei et~al\mbox{.}(2003)]%
        {blei2003latent}
\bibfield{author}{\bibinfo{person}{David~M Blei}, \bibinfo{person}{Andrew~Y Ng}, {and} \bibinfo{person}{Michael~I Jordan}.} \bibinfo{year}{2003}\natexlab{}.
\newblock \showarticletitle{Latent dirichlet allocation}.
\newblock \bibinfo{journal}{\emph{Journal of machine Learning research}} \bibinfo{volume}{3}, \bibinfo{number}{Jan} (\bibinfo{year}{2003}), \bibinfo{pages}{993--1022}.
\newblock


\bibitem[Buchan et~al\mbox{.}(2021)]%
        {buchan2021alignment}
\bibfield{author}{\bibinfo{person}{Jim Buchan}, \bibinfo{person}{Muneera Bano}, \bibinfo{person}{Didar Zowghi}, \bibinfo{person}{Stephen MacDonell}, {and} \bibinfo{person}{Amrita Shinde}.} \bibinfo{year}{2021}\natexlab{}.
\newblock \showarticletitle{Alignment of Stakeholder Expectations about User Involvement in Agile Software Development}.
\newblock \bibinfo{journal}{\emph{Empirical Software Engineering}} \bibinfo{volume}{26}, \bibinfo{number}{3} (\bibinfo{year}{2021}), \bibinfo{pages}{1275--1312}.
\newblock


\bibitem[Chang et~al\mbox{.}(2025)]%
        {chang2025lita}
\bibfield{author}{\bibinfo{person}{Chia-Hsuan Chang}, \bibinfo{person}{Jui-Tse Tsai}, \bibinfo{person}{Yi-Hang Tsai}, {and} \bibinfo{person}{San-Yih Hwang}.} \bibinfo{year}{2025}\natexlab{}.
\newblock \showarticletitle{LITA: An Efficient LLM-Assisted Iterative Topic Augmentation Framework}. In \bibinfo{booktitle}{\emph{Pacific-Asia Conference on Knowledge Discovery and Data Mining}}. Springer, \bibinfo{pages}{449--460}.
\newblock


\bibitem[Chang et~al\mbox{.}(2009)]%
        {chang2009reading}
\bibfield{author}{\bibinfo{person}{Jonathan Chang}, \bibinfo{person}{Sean Gerrish}, \bibinfo{person}{Chong Wang}, \bibinfo{person}{Jordan Boyd-Graber}, {and} \bibinfo{person}{David Blei}.} \bibinfo{year}{2009}\natexlab{}.
\newblock \showarticletitle{Reading tea leaves: How humans interpret topic models}.
\newblock \bibinfo{journal}{\emph{Advances in neural information processing systems}}  \bibinfo{volume}{22} (\bibinfo{year}{2009}).
\newblock


\bibitem[Creswell and Poth(2016)]%
        {creswell2016qualitative}
\bibfield{author}{\bibinfo{person}{John~W Creswell} {and} \bibinfo{person}{Cheryl~N Poth}.} \bibinfo{year}{2016}\natexlab{}.
\newblock \bibinfo{booktitle}{\emph{Qualitative inquiry and research design: Choosing among five approaches}}.
\newblock \bibinfo{publisher}{Sage publications}.
\newblock


\bibitem[Dong et~al\mbox{.}(2024a)]%
        {dong2024user}
\bibfield{author}{\bibinfo{person}{Jiong Dong}, \bibinfo{person}{Kaoru Ota}, {and} \bibinfo{person}{Mianxiong Dong}.} \bibinfo{year}{2024}\natexlab{a}.
\newblock \showarticletitle{User Experience of Different Groups in Social VR Applications: An Empirical Study Based on User Reviews}.
\newblock \bibinfo{journal}{\emph{IEEE Transactions on Computational Social Systems}} (\bibinfo{year}{2024}).
\newblock


\bibitem[Dong et~al\mbox{.}(2024b)]%
        {dong2024points}
\bibfield{author}{\bibinfo{person}{Jiong Dong}, \bibinfo{person}{Kaoru Ota}, {and} \bibinfo{person}{Mianxiong Dong}.} \bibinfo{year}{2024}\natexlab{b}.
\newblock \showarticletitle{What Are the Points of Concern for Players about VR Games: An Empirical Study based on User Reviews in Different Languages}.
\newblock \bibinfo{journal}{\emph{ACM Games: Research and Practice}} \bibinfo{volume}{2}, \bibinfo{number}{4} (\bibinfo{year}{2024}), \bibinfo{pages}{1--18}.
\newblock


\bibitem[Egger and Yu(2022)]%
        {egger2022topic}
\bibfield{author}{\bibinfo{person}{Roman Egger} {and} \bibinfo{person}{Joanne Yu}.} \bibinfo{year}{2022}\natexlab{}.
\newblock \showarticletitle{A topic modeling comparison between lda, nmf, top2vec, and bertopic to demystify twitter posts}.
\newblock \bibinfo{journal}{\emph{Frontiers in sociology}}  \bibinfo{volume}{7} (\bibinfo{year}{2022}), \bibinfo{pages}{886498}.
\newblock


\bibitem[Epp et~al\mbox{.}(2021)]%
        {epp2021empirical}
\bibfield{author}{\bibinfo{person}{Rain Epp}, \bibinfo{person}{Dayi Lin}, {and} \bibinfo{person}{Cor-Paul Bezemer}.} \bibinfo{year}{2021}\natexlab{}.
\newblock \showarticletitle{An empirical study of trends of popular virtual reality games and their complaints}.
\newblock \bibinfo{journal}{\emph{IEEE Transactions on Games}} \bibinfo{volume}{13}, \bibinfo{number}{3} (\bibinfo{year}{2021}), \bibinfo{pages}{275--286}.
\newblock


\bibitem[Gavgani et~al\mbox{.}(2017)]%
        {gavgani2017profiling}
\bibfield{author}{\bibinfo{person}{Alireza~Mazloumi Gavgani}, \bibinfo{person}{Keith~V Nesbitt}, \bibinfo{person}{Karen~L Blackmore}, {and} \bibinfo{person}{Eugene Nalivaiko}.} \bibinfo{year}{2017}\natexlab{}.
\newblock \showarticletitle{Profiling subjective symptoms and autonomic changes associated with cybersickness}.
\newblock \bibinfo{journal}{\emph{Autonomic Neuroscience}}  \bibinfo{volume}{203} (\bibinfo{year}{2017}), \bibinfo{pages}{41--50}.
\newblock


\bibitem[Grootendorst(2022)]%
        {grootendorst2022bertopic}
\bibfield{author}{\bibinfo{person}{Maarten Grootendorst}.} \bibinfo{year}{2022}\natexlab{}.
\newblock \showarticletitle{BERTopic: Neural topic modeling with a class-based TF-IDF procedure}.
\newblock \bibinfo{journal}{\emph{arXiv preprint arXiv:2203.05794}} (\bibinfo{year}{2022}).
\newblock


\bibitem[Han et~al\mbox{.}(2020)]%
        {han2020programmers}
\bibfield{author}{\bibinfo{person}{Junxiao Han}, \bibinfo{person}{Emad Shihab}, \bibinfo{person}{Zhiyuan Wan}, \bibinfo{person}{Shuiguang Deng}, {and} \bibinfo{person}{Xin Xia}.} \bibinfo{year}{2020}\natexlab{}.
\newblock \showarticletitle{What do programmers discuss about deep learning frameworks}.
\newblock \bibinfo{journal}{\emph{Empirical Software Engineering}}  \bibinfo{volume}{25} (\bibinfo{year}{2020}), \bibinfo{pages}{2694--2747}.
\newblock


\bibitem[Hardeniya et~al\mbox{.}(2016)]%
        {hardeniya2016natural}
\bibfield{author}{\bibinfo{person}{Nitin Hardeniya}, \bibinfo{person}{Jacob Perkins}, \bibinfo{person}{Deepti Chopra}, \bibinfo{person}{Nisheeth Joshi}, {and} \bibinfo{person}{Iti Mathur}.} \bibinfo{year}{2016}\natexlab{}.
\newblock \bibinfo{booktitle}{\emph{Natural language processing: python and NLTK}}.
\newblock \bibinfo{publisher}{Packt Publishing Ltd}.
\newblock


\bibitem[Hasan et~al\mbox{.}(2021)]%
        {hasan2021survey}
\bibfield{author}{\bibinfo{person}{Khalid Hasan}, \bibinfo{person}{Partho Chakraborty}, \bibinfo{person}{Rifat Shahriyar}, \bibinfo{person}{Anindya Iqbal}, {and} \bibinfo{person}{Gias Uddin}.} \bibinfo{year}{2021}\natexlab{}.
\newblock \showarticletitle{A Survey-Based Qualitative Study to Characterize Expectations of Software Developers from Five Stakeholders}.
\newblock \bibinfo{journal}{\emph{ACM Symposium on Application Performance Engineering}} (\bibinfo{year}{2021}).
\newblock


\bibitem[Hassan et~al\mbox{.}(2018)]%
        {hassan2018dialogue}
\bibfield{author}{\bibinfo{person}{Safwat Hassan}, \bibinfo{person}{Wei~Wang Shang}, \bibinfo{person}{Chakkrit Tantithamthavorn}, \bibinfo{person}{Cor-Paul Bezemer}, {and} \bibinfo{person}{Ahmed~E. Hassan}.} \bibinfo{year}{2018}\natexlab{}.
\newblock \showarticletitle{Studying the dialogue between users and developers of free apps in the Google Play store}.
\newblock \bibinfo{journal}{\emph{Empirical Software Engineering}} \bibinfo{volume}{23}, \bibinfo{number}{3} (\bibinfo{year}{2018}), \bibinfo{pages}{1275--1312}.
\newblock


\bibitem[{HTC Vive Developer Forums}(2024)]%
        {HTCDeveloper2024}
\bibfield{author}{\bibinfo{person}{{HTC Vive Developer Forums}}.} \bibinfo{year}{2024}\natexlab{}.
\newblock \bibinfo{title}{HTC Vive Developer Forums}.
\newblock \bibinfo{howpublished}{Retrieved Jul 2024 from \url{https://forum.htc.com/forum/24-vive-developer-forums/}}.
\newblock


\bibitem[Kamienski and Bezemer(2021)]%
        {kamienski2021empirical}
\bibfield{author}{\bibinfo{person}{Arthur Kamienski} {and} \bibinfo{person}{Cor-Paul Bezemer}.} \bibinfo{year}{2021}\natexlab{}.
\newblock \showarticletitle{An empirical study of Q\&A websites for game developers}.
\newblock \bibinfo{journal}{\emph{Empirical Software Engineering}} \bibinfo{volume}{26}, \bibinfo{number}{6} (\bibinfo{year}{2021}), \bibinfo{pages}{115}.
\newblock


\bibitem[Kapoor et~al\mbox{.}(2024)]%
        {kapoor2024qualitative}
\bibfield{author}{\bibinfo{person}{Satya Kapoor}, \bibinfo{person}{Alex Gil}, \bibinfo{person}{Sreyoshi Bhaduri}, \bibinfo{person}{Anshul Mittal}, {and} \bibinfo{person}{Rutu Mulkar}.} \bibinfo{year}{2024}\natexlab{}.
\newblock \showarticletitle{Qualitative insights tool (qualit): Llm enhanced topic modeling}.
\newblock \bibinfo{journal}{\emph{arXiv preprint arXiv:2409.15626}} (\bibinfo{year}{2024}).
\newblock


\bibitem[Karre et~al\mbox{.}(2019)]%
        {karre2019vr_practices}
\bibfield{author}{\bibinfo{person}{Sai~Anirudh Karre}, \bibinfo{person}{Neeraj Mathur}, {and} \bibinfo{person}{Y.~Raghu Reddy}.} \bibinfo{year}{2019}\natexlab{}.
\newblock \showarticletitle{Is Virtual Reality Product Development Different? An Empirical Study on VR Product Development Practices}. In \bibinfo{booktitle}{\emph{ISEC '19}}. \bibinfo{pages}{1--11}.
\newblock


\bibitem[Khalid et~al\mbox{.}(2015)]%
        {Khalid}
\bibfield{author}{\bibinfo{person}{Hammad Khalid}, \bibinfo{person}{Emad Shihab}, \bibinfo{person}{Meiyappan Nagappan}, {and} \bibinfo{person}{Ahmed~E. Hassan}.} \bibinfo{year}{2015}\natexlab{}.
\newblock \showarticletitle{What Do Mobile App Users Complain About?}
\newblock \bibinfo{journal}{\emph{IEEE Software}} \bibinfo{volume}{32}, \bibinfo{number}{3} (\bibinfo{year}{2015}), \bibinfo{pages}{70--77}.
\newblock
\href{https://doi.org/10.1109/MS.2014.50}{doi:\nolinkurl{10.1109/MS.2014.50}}


\bibitem[Kozlowski et~al\mbox{.}(2024)]%
        {kozlowski2024generative}
\bibfield{author}{\bibinfo{person}{Diego Kozlowski}, \bibinfo{person}{Carolina Pradier}, {and} \bibinfo{person}{Pierre Benz}.} \bibinfo{year}{2024}\natexlab{}.
\newblock \showarticletitle{Generative AI for automatic topic labelling}.
\newblock \bibinfo{journal}{\emph{arXiv preprint arXiv:2408.07003}} (\bibinfo{year}{2024}).
\newblock


\bibitem[Lenberg et~al\mbox{.}(2018)]%
        {lenberg2018misaligned}
\bibfield{author}{\bibinfo{person}{Per Lenberg}, \bibinfo{person}{Robert Feldt}, {and} \bibinfo{person}{Lars Göran~Wallgren Tengberg}.} \bibinfo{year}{2018}\natexlab{}.
\newblock \showarticletitle{Misaligned Values in Software Engineering Organizations}.
\newblock \bibinfo{journal}{\emph{arXiv preprint arXiv:1810.06104}} (\bibinfo{year}{2018}).
\newblock


\bibitem[Li et~al\mbox{.}(2023a)]%
        {li2023taggpt}
\bibfield{author}{\bibinfo{person}{Chen Li}, \bibinfo{person}{Yixiao Ge}, \bibinfo{person}{Jiayong Mao}, \bibinfo{person}{Dian Li}, {and} \bibinfo{person}{Ying Shan}.} \bibinfo{year}{2023}\natexlab{a}.
\newblock \showarticletitle{Taggpt: Large language models are zero-shot multimodal taggers}.
\newblock \bibinfo{journal}{\emph{arXiv preprint arXiv:2304.03022}} (\bibinfo{year}{2023}).
\newblock


\bibitem[Li et~al\mbox{.}(2023b)]%
        {li2023towards_quality}
\bibfield{author}{\bibinfo{person}{Shuqing Li}, \bibinfo{person}{Lili Wei}, \bibinfo{person}{Yepang Liu}, \bibinfo{person}{Cuiyun Gao}, \bibinfo{person}{Shing-Chi Cheung}, {and} \bibinfo{person}{Michael~R. Lyu}.} \bibinfo{year}{2023}\natexlab{b}.
\newblock \showarticletitle{Towards Modeling Software Quality of Virtual Reality Applications from Users' Perspectives}.
\newblock \bibinfo{journal}{\emph{arXiv preprint arXiv:2308.06783}} (\bibinfo{year}{2023}).
\newblock


\bibitem[Lin et~al\mbox{.}(2019)]%
        {lin2019empirical}
\bibfield{author}{\bibinfo{person}{Dayi Lin}, \bibinfo{person}{Cor-Paul Bezemer}, \bibinfo{person}{Ying Zou}, {and} \bibinfo{person}{Ahmed~E Hassan}.} \bibinfo{year}{2019}\natexlab{}.
\newblock \showarticletitle{An empirical study of game reviews on the Steam platform}.
\newblock \bibinfo{journal}{\emph{Empirical Software Engineering}}  \bibinfo{volume}{24} (\bibinfo{year}{2019}), \bibinfo{pages}{170--207}.
\newblock


\bibitem[Lu et~al\mbox{.}(2024)]%
        {lu2024empirical}
\bibfield{author}{\bibinfo{person}{Yijun Lu}, \bibinfo{person}{Kaoru Ota}, {and} \bibinfo{person}{Mianxiong Dong}.} \bibinfo{year}{2024}\natexlab{}.
\newblock \showarticletitle{An Empirical Study of VR Head-Mounted Displays Based on VR Games Reviews}.
\newblock \bibinfo{journal}{\emph{ACM Games: Research and Practice}} \bibinfo{volume}{2}, \bibinfo{number}{3} (\bibinfo{year}{2024}), \bibinfo{pages}{1--20}.
\newblock


\bibitem[Mauerer et~al\mbox{.}(2021)]%
        {mauerer2021socio}
\bibfield{author}{\bibinfo{person}{Wolfgang Mauerer}, \bibinfo{person}{Mitchell Joblin}, \bibinfo{person}{Damian Tamburri}, \bibinfo{person}{Carlos Paradis}, \bibinfo{person}{Rick Kazman}, {and} \bibinfo{person}{Sven Apel}.} \bibinfo{year}{2021}\natexlab{}.
\newblock \showarticletitle{In Search of Socio-Technical Congruence: A Large-Scale Longitudinal Study}.
\newblock \bibinfo{journal}{\emph{arXiv preprint arXiv:2105.08198}} (\bibinfo{year}{2021}).
\newblock


\bibitem[McKinney et~al\mbox{.}(2011)]%
        {mckinney2011pandas}
\bibfield{author}{\bibinfo{person}{Wes McKinney} {et~al\mbox{.}}} \bibinfo{year}{2011}\natexlab{}.
\newblock \showarticletitle{pandas: a foundational Python library for data analysis and statistics}.
\newblock \bibinfo{journal}{\emph{Python for high performance and scientific computing}} \bibinfo{volume}{14}, \bibinfo{number}{9} (\bibinfo{year}{2011}), \bibinfo{pages}{1--9}.
\newblock


\bibitem[{Meta}(2024)]%
        {Meta2024}
\bibfield{author}{\bibinfo{person}{{Meta}}.} \bibinfo{year}{2024}\natexlab{}.
\newblock \bibinfo{title}{Meta Official Website}.
\newblock \bibinfo{howpublished}{Retrieved Jul 2024 from \url{https://www.meta.com/}}.
\newblock


\bibitem[{Meta Community Forums}(2024)]%
        {MetaCommunity2024}
\bibfield{author}{\bibinfo{person}{{Meta Community Forums}}.} \bibinfo{year}{2024}\natexlab{}.
\newblock \bibinfo{title}{Meta Community Forums}.
\newblock \bibinfo{howpublished}{Retrieved Jul 2024 from \url{https://communityforums.atmeta.com/}}.
\newblock


\bibitem[Permana et~al\mbox{.}(2021)]%
        {permana2021stemming}
\bibfield{author}{\bibinfo{person}{Yudi Permana}, \bibinfo{person}{Arvita Emarilis}, {et~al\mbox{.}}} \bibinfo{year}{2021}\natexlab{}.
\newblock \showarticletitle{Stemming analysis indonesian language news text with Porter algorithm}. In \bibinfo{booktitle}{\emph{Journal of Physics: Conference Series}}, Vol.~\bibinfo{volume}{1845}. IOP Publishing, \bibinfo{publisher}{IOP Publishing}, \bibinfo{pages}{012019}.
\newblock


\bibitem[Pham et~al\mbox{.}(2023)]%
        {pham2023topicgpt}
\bibfield{author}{\bibinfo{person}{Chau~Minh Pham}, \bibinfo{person}{Alexander Hoyle}, \bibinfo{person}{Simeng Sun}, \bibinfo{person}{Philip Resnik}, {and} \bibinfo{person}{Mohit Iyyer}.} \bibinfo{year}{2023}\natexlab{}.
\newblock \showarticletitle{Topicgpt: A prompt-based topic modeling framework}.
\newblock \bibinfo{journal}{\emph{arXiv preprint arXiv:2311.01449}} (\bibinfo{year}{2023}).
\newblock


\bibitem[Ramage et~al\mbox{.}(2011)]%
        {ramage2011partially}
\bibfield{author}{\bibinfo{person}{Daniel Ramage}, \bibinfo{person}{Christopher~D Manning}, {and} \bibinfo{person}{Susan Dumais}.} \bibinfo{year}{2011}\natexlab{}.
\newblock \showarticletitle{Partially labeled topic models for interpretable text mining}. In \bibinfo{booktitle}{\emph{Proceedings of the 17th ACM SIGKDD international conference on Knowledge discovery and data mining}}. \bibinfo{pages}{457--465}.
\newblock


\bibitem[Reimers(2019)]%
        {reimers2019sentence}
\bibfield{author}{\bibinfo{person}{N Reimers}.} \bibinfo{year}{2019}\natexlab{}.
\newblock \showarticletitle{Sentence-BERT: Sentence Embeddings using Siamese BERT-Networks}.
\newblock \bibinfo{journal}{\emph{arXiv preprint arXiv:1908.10084}} (\bibinfo{year}{2019}).
\newblock


\bibitem[Rodr{\'\i}guez-P{\'e}rez et~al\mbox{.}(2021)]%
        {rodriguez2021perceived}
\bibfield{author}{\bibinfo{person}{Gema Rodr{\'\i}guez-P{\'e}rez}, \bibinfo{person}{Reza Nadri}, {and} \bibinfo{person}{Meiyappan Nagappan}.} \bibinfo{year}{2021}\natexlab{}.
\newblock \showarticletitle{Perceived diversity in software engineering: a systematic literature review}.
\newblock \bibinfo{journal}{\emph{Empirical Software Engineering}} \bibinfo{volume}{26}, \bibinfo{number}{5} (\bibinfo{year}{2021}), \bibinfo{pages}{102}.
\newblock


\bibitem[Rzig et~al\mbox{.}(2022)]%
        {rzig2022vr_testing}
\bibfield{author}{\bibinfo{person}{Dhia~Elhaq Rzig}, \bibinfo{person}{Nafees Iqbal}, \bibinfo{person}{Isabella Attisano}, \bibinfo{person}{Xue Qin}, {and} \bibinfo{person}{Foyzul Hassan}.} \bibinfo{year}{2022}\natexlab{}.
\newblock \showarticletitle{Characterizing Virtual Reality Software Testing}.
\newblock \bibinfo{journal}{\emph{arXiv preprint arXiv:2211.01992}} (\bibinfo{year}{2022}).
\newblock


\bibitem[Singh and O'Hagan(2024)]%
        {singh2024socialvr}
\bibfield{author}{\bibinfo{person}{Angelo Singh} {and} \bibinfo{person}{Joseph O'Hagan}.} \bibinfo{year}{2024}\natexlab{}.
\newblock \showarticletitle{Exploring Topic Modelling of User Reviews as a Monitoring Mechanism for Emergent Issues Within Social VR Communities}.
\newblock \bibinfo{journal}{\emph{arXiv preprint arXiv:2406.03989}} (\bibinfo{year}{2024}).
\newblock


\bibitem[Stahl(2020)]%
        {Stahl2020}
\bibfield{author}{\bibinfo{person}{Peter~M. Stahl}.} \bibinfo{year}{2020}\natexlab{}.
\newblock \bibinfo{title}{Language Detection}.
\newblock \bibinfo{howpublished}{Retrieved Jul 2, 2022 from \url{https://github.com/pemistahl/lingua}}.
\newblock


\bibitem[{Steam Community - App 250820}(2024)]%
        {SteamCommunityApp2024}
\bibfield{author}{\bibinfo{person}{{Steam Community - App 250820}}.} \bibinfo{year}{2024}\natexlab{}.
\newblock \bibinfo{title}{Steam Community Discussions for VR}.
\newblock \bibinfo{howpublished}{Retrieved Jul 2024 from \url{https://steamcommunity.com/app/250820}}.
\newblock


\bibitem[{Steam VR Store}(2024)]%
        {SteamVR2024}
\bibfield{author}{\bibinfo{person}{{Steam VR Store}}.} \bibinfo{year}{2024}\natexlab{}.
\newblock \bibinfo{title}{Steam VR Store Page}.
\newblock \bibinfo{howpublished}{Retrieved Jul 2024 from \url{https://store.steampowered.com/vr/}}.
\newblock


\bibitem[Team(2024)]%
        {market}
\bibfield{author}{\bibinfo{person}{VIVE Team}.} \bibinfo{year}{2024}\natexlab{}.
\newblock \bibinfo{title}{Virtual Reality Gaming Market Size, Share, Trends and Forecast by Segment, Device, Age Group, Type of Games, and Region, 2025-2033}.
\newblock \bibinfo{howpublished}{\url{https://www.imarcgroup.com/virtual-reality-gaming-market/}}.
\newblock
\newblock
\shownote{Accessed: 2025-07-23}.


\bibitem[Uddin et~al\mbox{.}(2021)]%
        {uddin2021empirical}
\bibfield{author}{\bibinfo{person}{Gias Uddin}, \bibinfo{person}{Fatima Sabir}, \bibinfo{person}{Yann-Ga{\"e}l Gu{\'e}h{\'e}neuc}, \bibinfo{person}{Omar Alam}, {and} \bibinfo{person}{Foutse Khomh}.} \bibinfo{year}{2021}\natexlab{}.
\newblock \showarticletitle{An empirical study of iot topics in iot developer discussions on stack overflow}.
\newblock \bibinfo{journal}{\emph{Empirical Software Engineering}}  \bibinfo{volume}{26} (\bibinfo{year}{2021}), \bibinfo{pages}{1--45}.
\newblock


\bibitem[Vergni et~al\mbox{.}(2021)]%
        {vergni2021evaluation}
\bibfield{author}{\bibinfo{person}{L Vergni}, \bibinfo{person}{F Todisco}, {and} \bibinfo{person}{B Di~Lena}.} \bibinfo{year}{2021}\natexlab{}.
\newblock \showarticletitle{Evaluation of the similarity between drought indices by correlation analysis and Cohen's Kappa test in a Mediterranean area}.
\newblock \bibinfo{journal}{\emph{Natural Hazards}} \bibinfo{volume}{108}, \bibinfo{number}{2} (\bibinfo{year}{2021}), \bibinfo{pages}{2187--2209}.
\newblock


\bibitem[{Viveport}(2024)]%
        {Viveport2024}
\bibfield{author}{\bibinfo{person}{{Viveport}}.} \bibinfo{year}{2024}\natexlab{}.
\newblock \bibinfo{title}{Viveport Official Website}.
\newblock \bibinfo{howpublished}{Retrieved Jul 2024 from \url{https://www.viveport.com/}}.
\newblock


\bibitem[Wan et~al\mbox{.}(2024)]%
        {wan2024tnt}
\bibfield{author}{\bibinfo{person}{Mengting Wan}, \bibinfo{person}{Tara Safavi}, \bibinfo{person}{Sujay~Kumar Jauhar}, \bibinfo{person}{Yujin Kim}, \bibinfo{person}{Scott Counts}, \bibinfo{person}{Jennifer Neville}, \bibinfo{person}{Siddharth Suri}, \bibinfo{person}{Chirag Shah}, \bibinfo{person}{Ryen~W White}, \bibinfo{person}{Longqi Yang}, {et~al\mbox{.}}} \bibinfo{year}{2024}\natexlab{}.
\newblock \showarticletitle{Tnt-llm: Text mining at scale with large language models}. In \bibinfo{booktitle}{\emph{Proceedings of the 30th ACM SIGKDD Conference on Knowledge Discovery and Data Mining}}. \bibinfo{pages}{5836--5847}.
\newblock


\bibitem[Zhang et~al\mbox{.}(2025)]%
        {zhang2025cvr_usability}
\bibfield{author}{\bibinfo{person}{Yawen Zhang}, \bibinfo{person}{Han Zhou}, \bibinfo{person}{Zhoumingju Jiang}, {and} \bibinfo{person}{Qinyuan Lei}.} \bibinfo{year}{2025}\natexlab{}.
\newblock \showarticletitle{Exploring Viewing Modalities in Cinematic Virtual Reality: A Systematic Review and Meta‑Analysis of User Experience}.
\newblock \bibinfo{journal}{\emph{Proceedings of the ACM on Human‑Computer Interaction}} (\bibinfo{year}{2025}).
\newblock


\end{thebibliography}

\appendix
\section{Appendix}
\label{Modeling}
This appendix provides a detailed description of the evaluation procedures and parameter selection process for the proposed HTModel.

\begin{figure}[t]
    \centering
    \begin{minipage}[t]{0.49\textwidth}
        \centering
        \includegraphics[width=\linewidth]{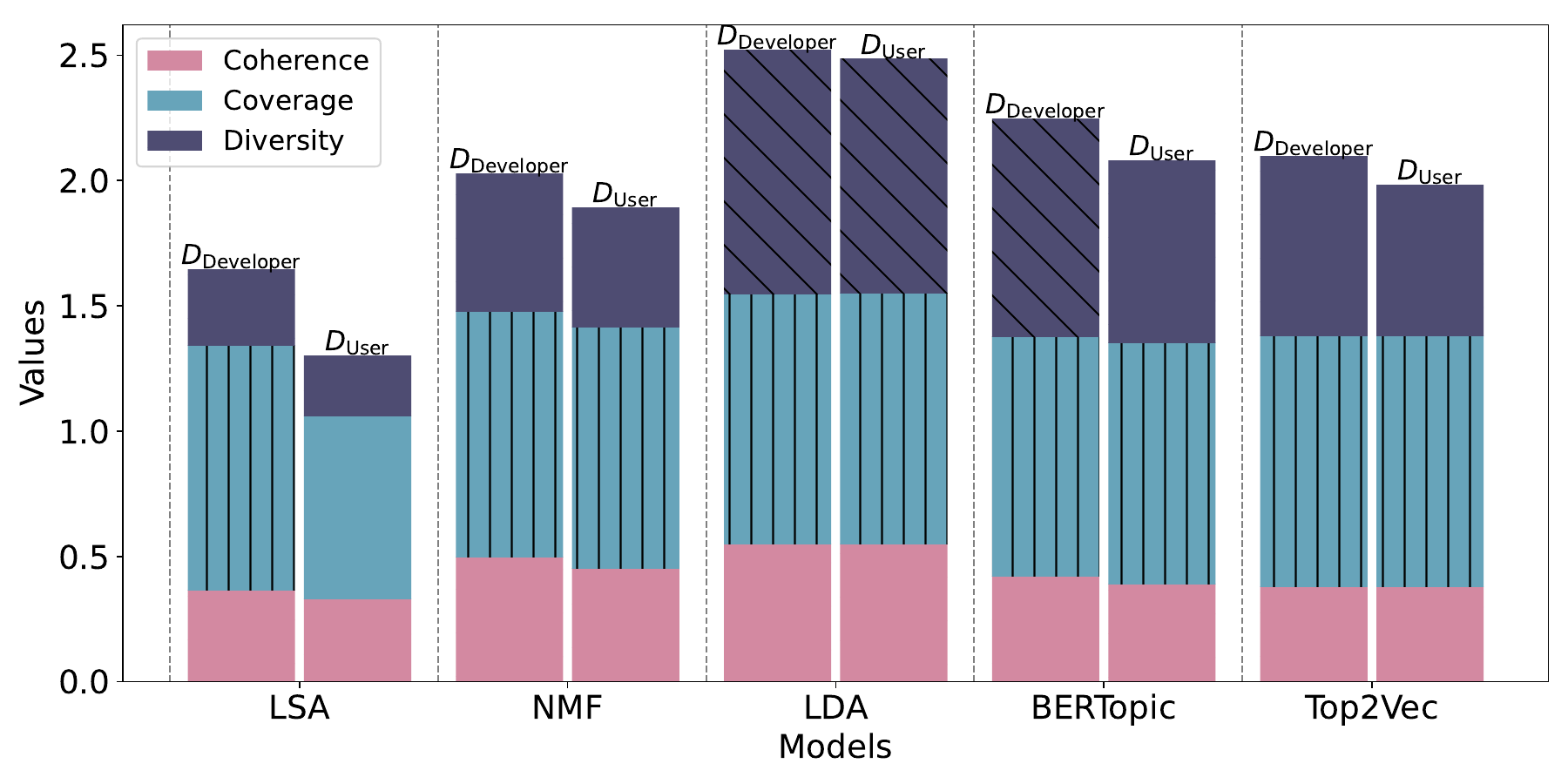}
        \caption*{(a) Quantitative evaluation of the topic models on the datasets $D_{developer}$ and $D_{user}$}
        \label{fig:topic_modeling_compare_quant}
    \end{minipage}
    \hfill
    \begin{minipage}[t]{0.49\textwidth}
        \centering
        \includegraphics[width=\linewidth]{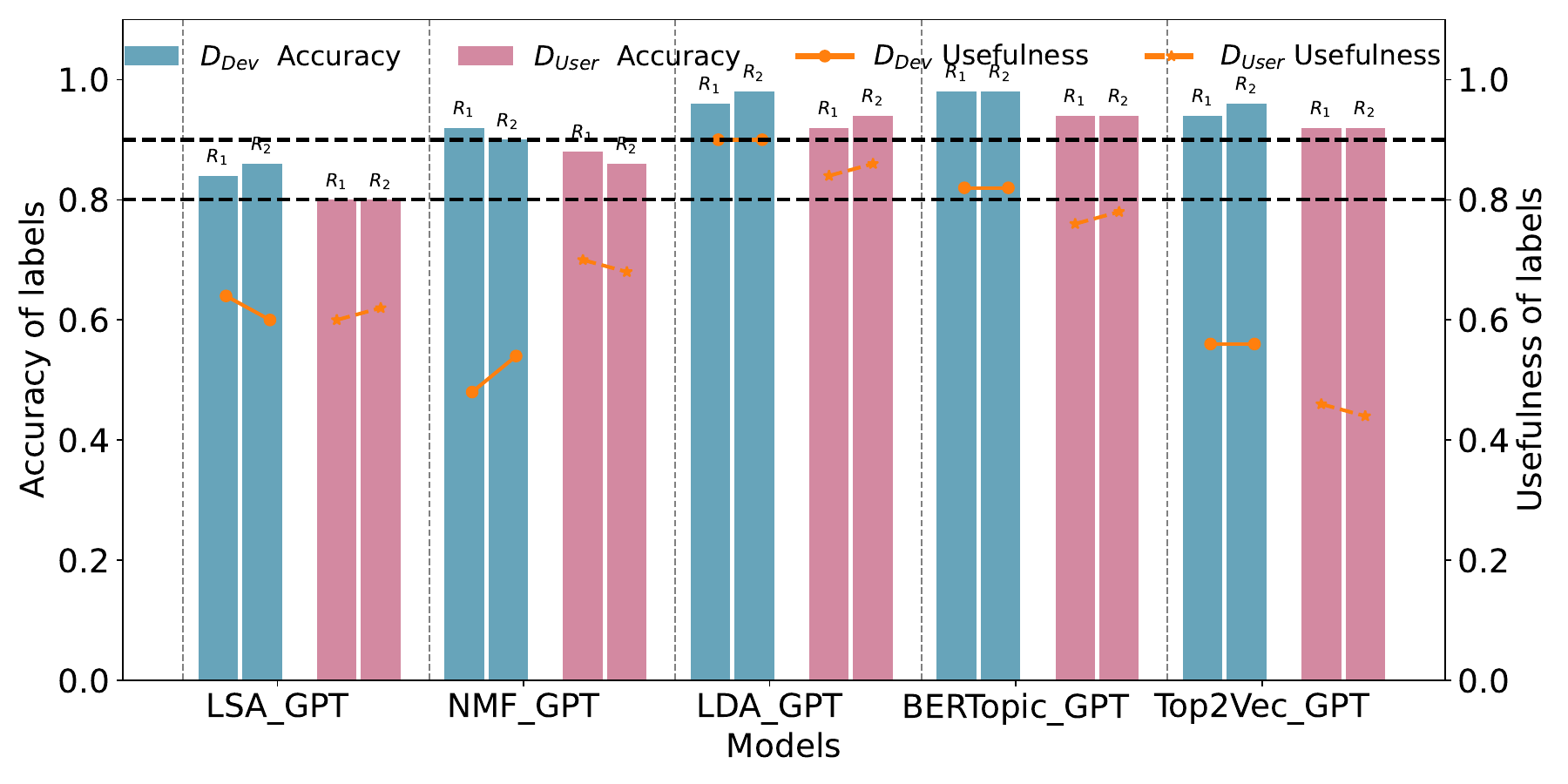}
        \caption*{(b) Qualitative evaluation of the topic models on the datasets $D_{developer}$ and $D_{user}$}
        \label{fig:topic_modeling_compare_qual}
    \end{minipage}
    \caption{
        Quantitative (a) and qualitative (b) evaluation of the topic models on the datasets $D_{developer}$ and $D_{user}$.
            \label{fig:topic_modeling_compare_all}
    }

\end{figure}

\subsection{Model Evaluation}
\label{Evaluation}

To evaluate the performance of the proposed HTModel for topic extraction in VR textual data, we design a comprehensive evaluation framework that combines both quantitative and qualitative evaluations. Here, the quantitative evaluation assesses the clustering effectiveness, and the qualitative evaluation examines the labeling quality. 

\subsubsection{Quantitative Evaluation}

We quantitatively evaluate the efficiency of conventional topic models in topic clustering using three key metrics, i.e., topic coherence, coverage, and diversity. The results are presented in Figure~\ref{fig:topic_modeling_compare_all} (a).


\textit{\textbf{Topic coherence}} is a critical metric used to assess the semantic coherence and meaningfulness of the topics generated by a topic model. We calculate topic coherence using the CoherenceModel class in Gensim, a widely used open-source Python library for topic modeling and document similarity analysis. The formula used to calculate topic coherence is expressed as follows:
\begin{equation}
    Coh_m= \frac{1}{T} \sum_{t=1}^{T} \sum_{i=1}^{|W_t|} \sum_{j=i+1}^{|W_t|} s(w_{ti}, w_{tj})
    \label{topic_coherence}
\end{equation}
where $m \in [1, 5]$ represents the model index, $t \in [1, 10]$ denotes the topic index, and $W_t$ denotes the set of words in topic $t$. Here, $w_{ti}$ and $w_{tj}$ are the $i$-th and $j$-th words in topic $W_t$, respectively, and $s(w_{ti}, w_{tj})$ measures the similarity between these words, which is typically computed using word co-occurrence statistics.

\textit{\textbf{Topic coverage}} quantifies the proportion of documents covered by the generated topics. Here, a higher coverage rate indicates that the topics are more representative of the dataset. The formula used to calculate topic coverage is expressed as follows:

\begin{equation}
    Cov_m = \frac{1}{D} \sum_{d=1}^{D} \mathbb{I}\left( \sum_{t=1}^T \mathbb{I}(P_{dt} > 0) > 0 \right)
    \label{topic_coverage}
\end{equation}
where $D$ denotes the total number of documents in the dataset. The topic coverage is determined by checking whether each document $d$ is associated with at least one topic, as indicated by the probability $P_{dt}$ of document $d$ belonging to topic $t$ being greater than $0$. The indicator function $\mathbb{I}(P_{dt} > 0)$ evaluates this condition.

\textit{\textbf{Topic diversity}} measures the distinctiveness of the topics generated by a model, where higher diversity values indicate that the topics cover a broader range of content and aspects in the dataset. The formula used to calculate topic diversity is expressed as follows:

\begin{equation}
    Div_m = \frac{1}{T} \sum_{t=1}^{T} \frac{| Unique(W_t) |}{| W_t |}
\end{equation}
where $|Unique(W_t)|$ represents the number of unique words in topic $t$, and $|W_t|$ is the total number of words in topic $t$. Here, a higher ratio of unique words to total words indicates greater topic diversity.



\subsubsection{Qualitative Evaluation}


We manually evaluate the labeling quality generated by GPT-4o using two key metrics: \textit{label accuracy} and \textit{label usefulness}. While our evaluation procedure is inspired by methodologies from previous studies \cite{wan2024tnt, li2023taggpt, kozlowski2024generative}, both metrics are uniquely defined in this work to better reflect the specific requirements of our research. This section provides a comprehensive assessment of the labels produced by the five models, with the results visualized in Figure~\ref{fig:topic_modeling_compare_all} (b). The details of our evaluation process and metric definitions are provided below.

\textit{\textbf{Label accuracy}} measures how well the assigned label can be directly inferred from the keyword group  and is categorized into three levels:

\begin{align}
    Acc_t =
    \begin{cases} 
        1, & \text{Bad} \ (\textit{Unrelated or misleading relative to the keyword group.}), \\ 
        2, & \text{Acceptable} \ (\textit{Captures part of the meaning but lacks clarity or completeness.}), \\ 
        3, & \text{Perfect} \ (\textit{Clearly and comprehensively represents the keyword group.}).
    \end{cases}
\end{align}

\textit{\textbf{Label usefulness}} evaluates the research significance of the label and is classified into the following three levels.

\begin{align}
    Use_t =
    \begin{cases} 
        1, & \text{Low} \ (\textit{Does not contribute meaningful value or insight}), \\ 
        2, & \text{Good} \ (\textit{Provides some value or partial contribution to the research}), \\ 
        3, & \text{High} \ (\textit{Offers significant value and demonstrates clear innovation or research impact}).
    \end{cases}
\end{align}

\subsubsection{Evaluation Process}

The qualitative evaluation was independently conducted by two experts, both of whom hold PhDs in Software Engineering and currently serve as senior UX designers at leading VR companies. Prior to assigning scores, the experts discussed and standardized the evaluation criteria, clarifying VR-related terminology and the interpretation of each rating level to ensure consistent and objective assessments.

For each topic label generated by the LDA\_GPT, LSA\_GPT, NMF\_GPT, BERTopic\_GPT, and Top2Vec\_GPT models, both experts independently rated both \textit{accuracy} and \textit{usefulness}. For instance, in terms of accuracy, a label such as “Gameplay Complexity” assigned to a cluster containing keywords like “challenge, difficulty, progression, level, strategy” was rated as \textit{Perfect} (score 3), since it precisely captures the semantic content of the cluster. In contrast, a label like “Miscellaneous” applied to a diverse set of unrelated keywords was rated as \textit{Bad} (score 1), due to its lack of specificity and interpretability. Labels such as “Customization” that sufficiently reflected most but not all of the cluster’s meaning were scored as \textit{Acceptable} (score 2).

Similarly, for usefulness, a label such as “Technical Issues” was considered \textit{Low} (score 1) as it is too broad for actionable development, while “Audio and Visual” was considered \textit{Good} (score 2), offering partial guidance but lacking specificity. In contrast, labels such as “Privacy \& Security” were assigned \textit{High} (score 3), as they are immediately actionable for design or engineering decisions.

To assess the consistency of the evaluation results, we employ \textit{Cohen’s Kappa coefficient} \cite{vergni2021evaluation}, which measures inter-rater reliability on a scale of [0, 1], where higher values indicate stronger agreement. The results for dataset $D_{developer}$ are $K_{LDA} = 0.83$, $K_{BERTopic} = 0.80$, $K_{Top2Vec} = 0.85$, $K_{LSA} = 0.84$, and $K_{NMF} = 0.87$. In addition, the results for dataset $D_{user}$ are $K_{LDA} = 0.81$, $K_{BERTopic} = 0.82$, $K_{Top2Vec} = 0.84$, $K_{LSA} = 0.83$, and $K_{NMF} = 0.85$. These values indicate strong agreement between the raters.

To quantify the metrics, we calculate the proportion of labels with scores $\geq 2$ for both accuracy and usefulness, transforming them into a normalized scale of 0–1. Here, values closer to 1 indicate higher label quality, where 1 represents perfect accuracy or usefulness. This approach enables an objective comparison of the models' performance in generating meaningful and precise labels.

\begin{table}[t]
\centering
\caption{Comparison of Cost, Human Involvement, and Interpretability for Major Topic Modeling Pipelines}
\label{tab:cost_compare}
\begin{tabular}{lcccc}
\toprule
\textbf{Method} & \textbf{LLM Usage} & \textbf{Cost (1M docs)} & \textbf{Human Effort} & \textbf{Interpretability} \\
\midrule
LDA/LSA/NMF & \ding{53}  & - & High (manual labeling) & Low (keywords only) \\
BERTopic/Top2Vec & \ding{53}  & - & High (manual labeling) & Low (keywords only) \\
TopicGPT &  \checkmark & High (\$2,700--\$10,800) &  Low (validation)  &  High  \\
HTModel &  \checkmark & Low (\$1--\$10) & Low (validation) & High \\
\bottomrule
\end{tabular}
\end{table}

\begin{figure}[t]
    \centering
    \includegraphics[width=\columnwidth]{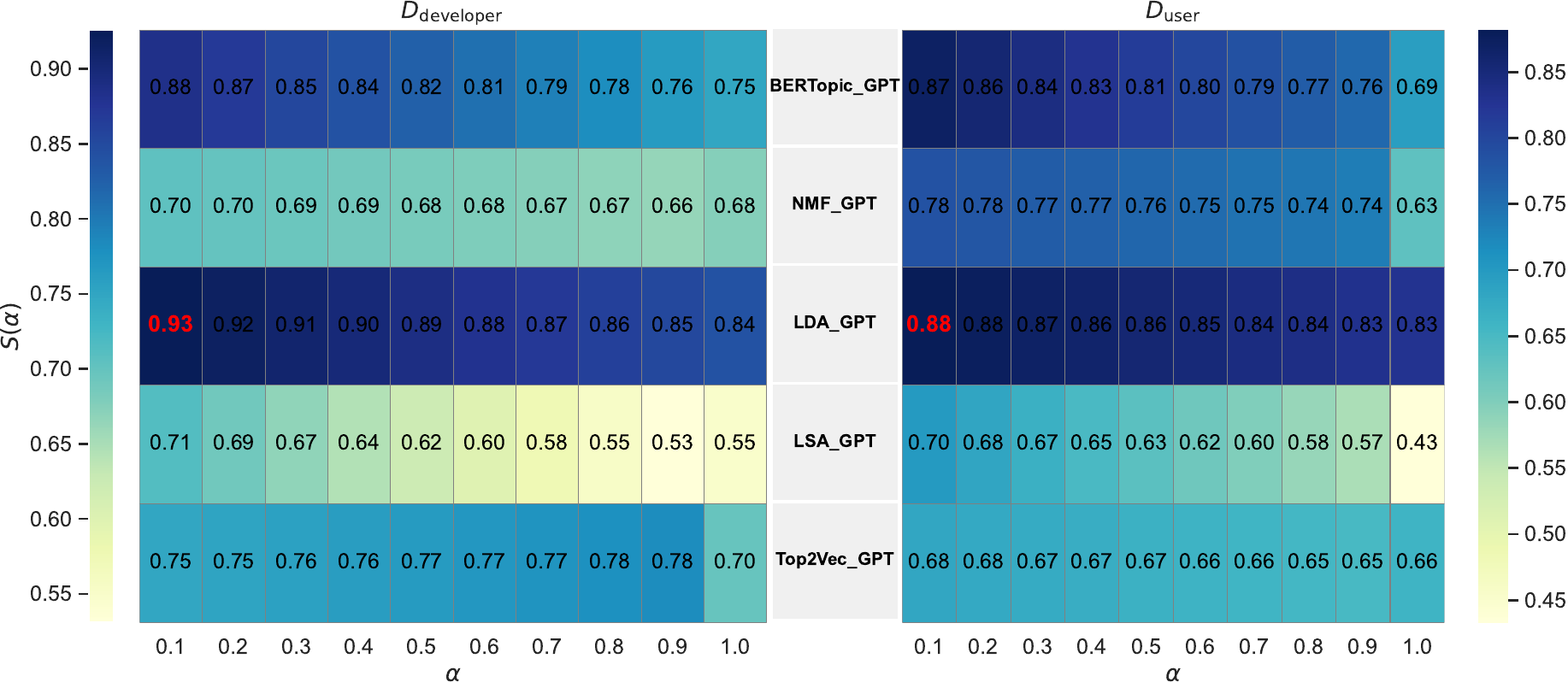}
    \caption{Distribution of Comprehensive Scores ($S$) for $D_{developer}$ and $D_{user}$ Across Topic Models and Weight Parameters ($\alpha$)}
    \label{Topic_Score}
\end{figure}

\subsubsection{Baseline Comparison}

Recent advances such as TopicGPT~\cite{pham2023topicgpt} demonstrate the potential of large language models (LLMs) for fully automated, highly interpretable topic modeling. Here, we compare the computational cost, human effort, and interpretability among traditional topic models (LDA, LSA, NMF, BERTopic, Top2Vec), end-to-end LLM-based frameworks (e.g., TopicGPT) and hybrid pipelines (traditional topic model clustering + LLM-based labeling, as in our HTModel). 

Table~\ref{tab:cost_compare} summarizes key differences. Traditional models are efficient and require almost no computational cost, but manual intervention is needed to interpret and label topics, which is labor-intensive and subjective. Hybrid approaches like HTModel automate topic extraction and use LLMs to label a limited set of topics, resulting in very low cost (e.g., labeling 50–100 topics via GPT-4o costs less than \$ 10 for over one million documents) and much greater interpretability, with minimal human involvement required for validation. In contrast, end-to-end LLM frameworks such as TopicGPT offer the highest interpretability, automatically generating natural language topics and assignments, but at much higher computational cost (e.g., \$88–\$155 for tens of thousands of documents), since LLMs are called repeatedly for generation, assignment, and refinement.


\subsection{Model Selection}
\label{Selection}

To support the subsequent multi-perspective empirical analysis, we design the following weighted scoring formula to select the optimal pipeline for topic modeling and label generation:
\begin{equation}
S_m = \alpha \cdot \frac{Coh_m+Cov_m+Div_m}{3} + (1 - \alpha) \cdot \frac{Q_{m}}{2} 
\end{equation}
where $S_m$ denotes the comprehensive model scores of the $m$-th model, and $\alpha \in [0, 1]$ is a weight parameter that balances the importance of quantitative and qualitative evaluation metrics. 
$Q_m = \frac{Q_{m1} + Q_{m2}}{2}$, where $Q_{m1} = \frac{1}{T} \sum_{t=1}^T (Acc_t + Use_t)$ denotes the scores assigned by the evaluators. 
A larger value of $\alpha$ emphasizes quantitative evaluation (i.e., topic coherence, diversity, and coverage), while a smaller $\alpha$ gives greater weight to qualitative evaluation (i.e., label accuracy and usefulness). This parameter can be flexibly adjusted according to the specific requirements of the task and application scenario.


We conduct a systematic sensitivity analysis of $\alpha$ in the range from 0.1 to 1.0. As shown in Figure~\ref{Topic_Score}, the LDA-GPT model achieves its highest comprehensive score $S$ when $\alpha=0.1$ ($S_{LDA-GPT}=0.93$ on $D_{developer}$ and $S_{LDA-GPT}=0.88$ on $D_{user}$). Therefore, we adopt $\alpha=0.1$ as the default weight setting for subsequent experiments.

\begin{figure}[t]
    \centering
    \begin{minipage}[t]{0.49\textwidth}
        \centering

                \includegraphics[height=4cm]{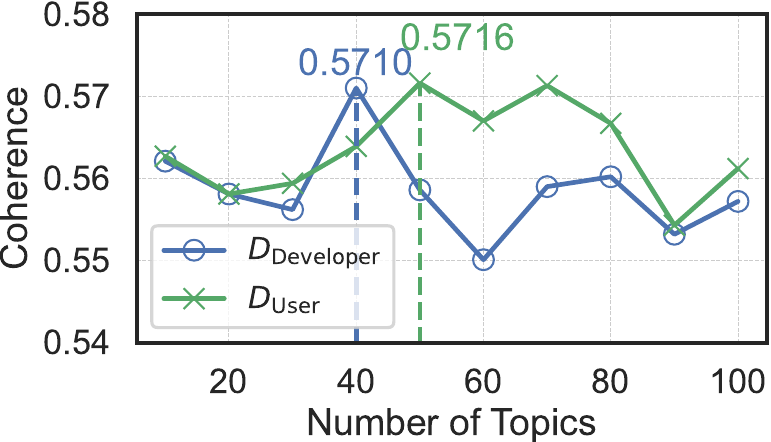}
        \caption*{(a) Topics}
        \label{topic}
    \end{minipage}
    \begin{minipage}[t]{0.49\textwidth}
        \centering

        \includegraphics[height=3.9cm]{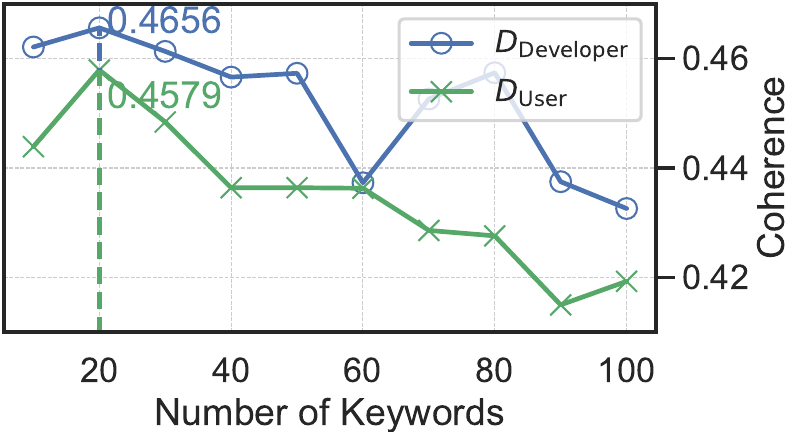}
        \caption*{(b) Keywords}
         \label{kewods}
    \end{minipage}
\caption{
Coherence scores versus the number of topics (a) and keywords (b) for $D_{developer}$ and $D_{user}$.
}

    \label{LDA_topic_modeling}
\end{figure}

After determining the optimal model and weight parameter, we further tune two key hyperparameters of the LDA model: the number of topics ($T$) and the number of keywords per topic ($K$). Specifically, for both the developer and user datasets, we perform a sensitivity analysis by varying the number of topics $T$ over $\{10, 20, 30, \ldots, 100\}$ and compute the topic coherence score for each configuration. The results indicate that the highest coherence is achieved at $T=40$ for the developer dataset and $T=50$ for the user dataset. As shown in  Figure~\ref{LDA_topic_modeling}(a). Thus, we select $T_{D_{\text{Developer}}}=40$ and $T_{D_{\text{User}}}=50$ as the optimal numbers of topics for subsequent analysis.

Building upon the optimal topic numbers, we further perform a sensitivity analysis on the number of keywords per topic $K$ (i.e., $K=10, 20, 30, \ldots, 100$). The results show that the topic coherence score reaches its maximum when $K=20$ for both datasets (see Figure~\ref{LDA_topic_modeling}(b)). Therefore, we set the number of keywords per topic to 20, striking a balance between the richness of topic description and the semantic cohesion of the resulting topics.

\begin{tcolorbox}[
                    colback=gray!8, 
                    colframe=black, 
                    width=\linewidth, 
                    arc=1mm, 
                    auto outer arc,
                    boxrule=0.8pt,
                    boxsep=1pt, 
                    left=2pt, 
                    right=2pt, 
                    top=2pt, 
                    bottom=2pt 
                 ]               
\textbf{Finding 8:} The proposed HTModel exhibits strong and consistent performance across different model combinations. All hybrid models achieve over 80\% labeling accuracy, with LDA\_GPT, BERTopic\_GPT, and Top2Vec\_GPT exceeding 90\%. Notably, only LDA\_GPT and BERTopic\_GPT surpass 80\% in labeling usefulness, with LDA\_GPT achieving the best overall performance on our multi-stakeholder dataset. Importantly, hybrid pipelines offer an effective trade-off between cost, automation, and interpretability, making them particularly well-suited for large-scale empirical studies. 

\end{tcolorbox}

\end{document}